\documentclass[a4paper,11pt]{article}
\usepackage{jheppub} 
\usepackage{amsfonts,amssymb,epsfig,amsmath,mathtools}
\usepackage{color}
\usepackage{comment}
\usepackage{url}
\usepackage{tikz}
\allowdisplaybreaks
\usepackage{youngtab}
\usepackage{mathrsfs}
\usepackage{graphicx}
\usepackage{subcaption}
\usepackage{hyperref}
\newcommand{\polp}{\epsilon^\parallel}
\newcommand{\polo}{\epsilon^\perp}

\newcommand{\be}{\begin{eqnarray}}
\newcommand{\ee}{\end{eqnarray}}

\newcommand{\bn}{\begin{enumerate}}
\newcommand{\en}{\end{enumerate}}

\newcommand{\Z}{\mathbb{Z}}
\newcommand{\tyng}{\tiny\yng}

\preprint{TIFR/TH/23-20}
\title{\boldmath 
\vspace*{-0.3cm}
Regge constraints on local four-point scattering amplitudes of massive particles with spin
\vspace{-0.2cm}
}

\author[a,1]{Subham Dutta Chowdhury,\note{subham@uchicago.edu}}
\author[b,2]{Vipul Kumar,\note{vipul.kumar@tifr.res.in}}
\author[c,3]{Suman Kundu,\note{suman.kundu@weizmann.ac.il}}
\author[b,4]{Asikur Rahaman.\note{asikur.rahaman@tifr.res.in}}
\affiliation[a]{Kadanoff Center for Theoretical Physics \& Enrico Fermi Institute,\\
University of Chicago, Illinois 60615, USA}
\affiliation[b]{Department of Theoretical Physics, \\Tata Institute of Fundamental Research, Homi Bhabha Rd, Mumbai 400005, India}
\affiliation[c]{Department of Particle Physics and Astrophysics, \\The Weizmann Institute of Science, Rehovot 76100, Israel.
 \vspace{-0.3cm}}

\abstract{In this work, we  classify all the possible local four-point couplings relevant for tree-level flat space $2 \rightarrow 2$ scattering of external massive particles of spin one and spin two which do not grow faster than $s^2$ at large $s$ and fixed t. This kinematic constraint on local growth of tree-level S-matrices is known as Classical Regge Growth criteria or CRG \cite{Chowdhury:2019kaq}. We first construct the spin one and spin two tree-level contact S-matrices as modules of polarisation tensors and momenta over the ring of polynomials generated by Mandelstam invariants.  We then consider a general scattering process where the external scattering particles are of different masses but of same spin and constrain this space to obtain a finite number of CRG allowed local Lagrangians. Our concrete results are primarily for $D\geq 8$ but the process outlined is easily generalised to lower dimensions to include low dimensional parity violating structures. The space of CRG allowed structures reduces when we specialise to identical scattering and restrict to parity even couplings in $D=4$. We show that tree-level scattering amplitudes involving exchange diagrams and contact terms in de Rham-Gabadadze-Tolley massive gravity (dRGT) violate CRG unless the parameters of the theory take special values. The CRG allowed S-matrices, in the context of large $N$ conformal field theories (CFTs), can also be interpreted as bulk $AdS$ counterterms consistent with Chaos bound. Our classified structures therefore can be thought of as ambiguities arising in the context of conformal field theory inversion formula for four point functions of unconserved spin one and spin two operators in large $N$ CFTs.    
}

\begin{document}
\maketitle
\flushbottom

\section{Introduction} \label{intro}

In interacting local quantum field theories, tree-level scattering amplitudes are generated by local Lagrangians, given schematically by the interactive part of the action, 
 \begin{equation}\label{lagintro}
     \mathcal{S}_{int}= \int d^D x~ \sum_i \gamma_i L_i,
 \end{equation}
 where $D$ denotes the spacetime dimensions,  $\gamma_i$ are dimensionful couplings and $L_i$ are various higher derivative local interaction Lagrangians of the fields, whose scattering process we want to study. The tree-level amplitudes exhibit two main analytic structures. It involves simple pole-type singularities in momenta, due to exchange of massless or massive particles, while the other one comprises only of polynomial terms, known as contact terms. The contact pieces are completely  analytic in momenta and polarisation tensors (corresponding to external spinning states). For $2 \to 2$ tree-level scattering processes which are of interest in this work the local Lagrangians can be at most quartic in the fields, but they can be of arbitrary high order in derivatives. 
 
 In \cite{Chowdhury:2019kaq, Chakraborty:2020rxf}, the authors proposed constraints on such Lagrangians which generate tree-level amplitudes for four photon and four graviton scattering. The criteria called Classical Regge Growth or CRG (which was subsequently proved in \cite{Chandorkar:2021viw} using flat space limit of $AdS/CFT$), states that four point tree-level scattering amplitudes should always grow slower than or equal to $s^2$ at large $s$ and fixed $t$ (where $s$ and $t$ are Mandelstam variables) for all values of the physical momenta and for all values of the polarisation tensors (if the external states have spin). This kinematic limit is also known as Regge limit of flat space scattering in literature. Because of the presence of dimensionful couplings associated with these Lagrangians, it is important to state the energy scales we are working in. Let us assume that we are working in a theory where there exists a hierarchy of scales, and our Regge limit probes the energy regimes between the IR and UV scales. An example where one encounters such a hierarchy is for example tree-level String theory. The dimensionful couplings that parametrize interactions can be suppressed by the string length while the UV scale is the parameterised Planck length. Another place where one encounters similar hierarchy of scales is Kaluza Klein reduction of Einstein gravity where the relevant lower scale would be set by the length of the compactification circle. For  the four point tree-level scattering amplitude $T(s,t)$, generated by the Lagrangian eq \eqref{lagintro}, CRG then states that, 

\begin{eqnarray}
    \lim_{s \rightarrow \infty,~ \textrm{fixed} ~t<0} T(s,t) \leq s^2 ,\qquad
\frac{1}{L_{IR}^2} \ll s\ll \frac{1}{L_{UV}^2}, 
\end{eqnarray}
where $L_{IR}$ is the string length for tree-level string theory or the compactification length for the Kaluza Klein theory and $L_{UV}$ is the Planck length $l_p$ for these examples.  Thus our thought experiment truly probes the tree-level processes in such theories, since loops in these examples would be suppressed by $l_p$. One of the striking results of this criteria was to rule out any finite polynomial modifications of Einstein Hilbert action in $D\leq 6$, upto quartic order\footnote{See also \cite{Camanho:2014apa} for constraints on modification to Einstein Hilbert action (upto cubic order) from causality principles.}. More recently the authors of \cite{Haring:2022cyf}, proved CRG directly in flat space for massive and massless scalar particles from flat space dispersion analysis. In particular, we refer the interested reader to section VI of \cite{Haring:2022cyf} for a proof of this bound on finite growth of scattering amplitudes. Note that CRG is different from the Froissart-Martin bound \cite{PhysRev.123.1053, Martin:1965jj} which is a bound on the full quantum amplitude (i.e including loops) of a quantum field theory with a mass gap (also generalised to massless theories in \cite{Caron-Huot:2021enk, Haring:2022cyf}). Froissart Martin bound is applicable for energies $s\gg \frac{1}{L_{UV}^2}$, and the growth exponent must be strictly less than two. Our present investigation encompasses generalising the work of \cite{Chowdhury:2019kaq} to constrain the space of contact term like amplitudes generated by four-point Lagrangians of massive spin one and spin two particles using CRG.

In order to classify the Lagrangians which generate such local tree-level amplitudes, we employ the approach adopted in \cite{Chowdhury:2019kaq}. Consider the $2 \rightarrow 2$ tree-level scattering of massive spinning particles of different masses generated by all possible contact term like interactions. The scattering amplitudes are functions of polarisation tensors describing the external spinning state and the Mandelstam invariants $s,t$ and $u$ satisfying $s+t+u= \sum_i m_i^2$. The relevant interaction Lagrangians generating these amplitudes are contact terms of arbitrary high derivative order, quartic in the spin one and spin two fields. Lagrangians which give rise to the set of tree-level amplitudes form an equivalence class - if two Lagrangians differ by equations of motion and total derivatives they give rise to the same scattering amplitude. This ensures that there is a one-to-one map between the space of contact scattering amplitudes and the equivalence class of local Lagrangians. Thus classifying contact scattering amplitudes is equivalent to writing down local Lagrangians.

Massive spinning polarisations have no gauge redundancy hence any Lorentz scalar polynomial of polarisation and momenta with the correct homogeneity in polarisation tensors is an allowed contact amplitude and therefore a local Lagrangian. It is useful to think of this infinite space of analytic scattering amplitudes as a module over the ring of
polynomials of Mandelstam invariants following \cite{Henning:2017fpj, Chowdhury:2019kaq}. The seemingly infinite set of amplitudes is then finitely generated. The generators of these modules, called ``primaries", are polynomials of scalar products of polarisations with momenta and polarisations with each other obeying correct homogeneity. The Mandelstam invariants then act as ``descendants" of these generators giving rise to amplitudes for arbitrarily high order in derivatives. Unlike the massless case, the module generators need not be only local Lagrangians of field strength $F_{\mu \nu}$  for spin one or Riemann tensor $R_{\mu\nu\alpha\beta}$ for spin two. Starting from zeroth order in derivatives, it is then straightforward to classify the space of such linearly independent polynomials order by order in derivatives. This is a finite task which truncates at four order in derivatives for the spin one and eight derivatives for the massive spin two case. If we denote the primaries as $g_i$ and the ring elements as $r_i$, the most general S-matrix is given by $\sum_i g_i \cdot r_i$. In this work we restrict our analysis to  $D\geq 8$ where this module is freely generated. 

We also generalise to identical scattering, which implies imposing $S_4$ invariance on the space of primaries. It is useful to think of  $S_4$ as a semi-direct product  $S_3\ltimes (\Z_2\times \Z_2)$. Thus, as explained in Appendix \ref{s3}, we can impose this symmetry in two steps. We can first impose $\Z_2\times \Z_2$ symmetry on the primaries, giving rise to the ``quasi-invariant" or ``local module" which generate the amplitude corresponding to identical scattering. In the second step, we impose the remnant $S_3$ on these local modules. The local modules transform in irreducible representations of $S_3$ and the S-matrix corresponding to identical scattering is obtained by singlet $S_3$ projection of the quasi-invariants and its descendants \cite{Chowdhury:2019kaq}. The complete classification of primaries and quasi-invariants for spin one and spin two are recorded in the ancillary \texttt{Mathematica} files
\textit{spinone\_regge.nb} and \textit{spintwo\_regge.nb}

In order to study Regge growth of these amplitudes, it is also necessary to classify the independent data in scattering. We provide a convenient parametrization of massive polarisation tensors in \S \ref{kinematics} (see eq \eqref{spinonepol} and eq \eqref{polpparaspinone}) utilizing the equations of motion of massive spinning fields and $SO(D-3)$ symmetry of the four point scattering amplitude. Since the scattering process takes place in a three plane defined by the momentum conservation equation $\sum^4_{i=1} p^\mu_i=0$, the scattering amplitude therefore must be left invariant by $SO(D-3)$ rotations orthogonal to the plane of scattering. The kinematics of the scattering process also enables us to count the number of primaries and quasi invariants as the number of $SO(D-3)$ singlets that can be constructed from product of representations of little group of massive spinning particles with definite symmetry. We find agreement with our explicit construction of the basis of primaries and quasi-invariants.

For non-identical scattering, a single Lagrangian can give rise to twenty four different scattering processes, due to permutations of incoming and outgoing particles. These different scattering processes translate to different $S_4$ transformations of our primaries, i.e the generator of the amplitudes. CRG then translates to the fact that the primaries must grow slower than $s^2$ at large $s$ and fixed $t$ for each of these scattering processes. We find that for spin one twenty six couplings are CRG allowed (see section \S\ref{ReggeSpin1}). Out of the twenty six structures there are two  structures at four derivatives, twenty one structures at two derivatives and three structures at zero derivatives. We have listed them explicitly in the ancillary file \textit{spinone\_regge.nb}. For spin two non-identical case, we found total five CRG allowed structures. Out of these five structures, four are two-derivative structures and there is one zero derivative structure. We have listed them explicitly in eq \eqref{crgallowedspin2nonid}. One can also consider descendants of our primaries which would correspond to constraints on higher derivative couplings but we find that generically the space of CRG allowed primaries and descendants can be very large for non-identical scattering and we don't record them in this work.

For identical case we consider both local modules and descendants. There are ten CRG allowed structures for spin one generated by the Lagrangians eq. \eqref{spin1crgalwdlag} 
\begin{equation}\label{introspinonelag}
\begin{split}
\tilde{\mathcal{L}}_{\partial^0, 1}^{\mathfrak{v}} &= (A_\mu A^\mu)^2\\
\tilde{\mathcal{L}}_{\partial^2, 1}^{\mathfrak{v}} &=\partial_\sigma A_\mu A^\mu \partial^\sigma A_\nu A^\nu, \qquad\tilde{\mathcal{L}}_{\partial^2, 2}^{\mathfrak{v}} = \tilde{F}_{\alpha\beta}\tilde{F}^{\alpha\beta} A_\nu A^\nu,\qquad \tilde{\mathcal{L}}_{\partial^2, 3}^{\mathfrak{v}}= A^\mu F_{\mu \nu} \tilde{F}^{\nu \alpha} A_\alpha\\
\tilde{\mathcal{L}}_{\partial^4, 1}^{\mathfrak{v}} &= F_{\alpha \beta}F^{\beta \gamma}F_{\gamma \delta }F^{\delta \alpha},\qquad \tilde{\mathcal{L}}_{\partial^4, 2}^{\mathfrak{v}} = (F_{\alpha\beta}F^{\alpha\beta})^2, \qquad \tilde{\mathcal{L}}_{\partial^4, 3}^{\mathfrak{v}}= F_{ab}\tilde{F}^{bc}F_{cd}\tilde{F}^{da},\\
\tilde{\mathcal{L}}_{\partial^4, 4}^{\mathfrak{v}} &=A_{\mu} \partial_\sigma A_\nu \partial^\mu \partial^\sigma A_\alpha \partial^\nu A^\alpha\\
\tilde{\mathcal{L}}_{\partial^6, 1}^{\mathfrak{v}} &= \partial_\sigma F_{\alpha\beta}F^{\beta \gamma }\partial^\sigma F_{\gamma \delta }F^{\delta \alpha}, \qquad \tilde{\mathcal{L}}_{\partial^6, 2}^{\mathfrak{v}} =  F_{\alpha\beta} \partial^\alpha F^{\gamma \delta}\partial^\beta F_{\delta \eta}F^{\eta \gamma},
\end{split}
\end{equation}   
where we have defined $F_{ab}= \partial_a A_b-\partial_b A_a$ and $\tilde{F}_{ab}= \partial_a A_b+\partial_b A_a$. $A_\mu$ denotes the massive spin one field. There are four CRG allowed structures for spin two generated by the local Lagrangians eq. \eqref{basisspin2id}. 
\begin{equation}\label{introspintwolag}
    \begin{split}
    \tilde{\mathcal{L}}_{\partial^0}^{\mathfrak{h}} &=  \,\, \delta_{[\alpha}^{\gamma}\delta_{\beta}^{\zeta}\delta_{\xi}^{\rho}\delta_{\delta]}^{\sigma}\,\,  h_\gamma^{\phantom{a}\alpha} \, h_\zeta^{\phantom{a}\beta}\,  h_\rho^{\phantom{a}\zeta}\,h_\sigma^{\phantom{a}\delta},\qquad
     \tilde{\mathcal{L}}_{\partial^2}^{\mathfrak{h}} =  \,\, \delta_{[\alpha}^{\gamma}\delta_{\beta}^{\zeta}\delta_{\xi}^{\rho}\delta_{\delta}^{\sigma}\delta_{\mu]}^{\nu}\,\, \partial_\nu \partial^\mu h_\gamma^{\phantom{a}\alpha} \, h_\zeta^{\phantom{a}\beta}\,  h_\rho^{\phantom{a}\zeta}\,h_\sigma^{\phantom{a}\delta}, \\
    \tilde{\mathcal{L}}_{\partial^4}^{\mathfrak{h}} &=  \,\, \delta_{[\alpha}^{\gamma}\delta_{\beta}^{\zeta}\delta_{\xi}^{\rho}\delta_{\delta}^{\sigma}\delta_{\mu}^{\nu}\delta_{\mu']}^{\nu'}\,\, \partial_\nu \partial^\mu h_\gamma^{\phantom{a}\alpha} \, \partial_{\nu'} \partial^{\mu'} h_\zeta^{\phantom{a}\beta}\,  h_\rho^{\phantom{a}\zeta}\,h_\sigma^{\phantom{a}\delta},\\
        \tilde{\mathcal{L}}_{\partial^6}^{\mathfrak{h}} &=    \,\, \delta_{[\alpha}^{\gamma}\delta_{\beta}^{\zeta}\delta_{\xi}^{\rho}\delta_{\delta}^{\sigma}\delta_{\mu}^{\nu}\delta_{\mu'}^{\nu'}\delta_{\mu'']}^{\nu''}\,\, \partial_\nu \partial^\mu h_\gamma^{\phantom{a}\alpha} \, \partial_{\nu'} \partial^{\mu'} h_\zeta^{\phantom{a}\beta}\,  \partial_{\nu''} \partial^{\mu''}h_\rho^{\phantom{a}\zeta}\,h_\sigma^{\phantom{a}\delta}, \\
    \end{split}
\end{equation}

where $h_{\mu\nu}$ is the massive spin two field and  $\delta_{[\alpha_1}^{\mu_1}\delta_{\alpha_2}^{\mu_2}\cdots\delta_{\alpha_n]}^{\mu_n}$ is concise expression for product of two Levi-Civita tensors $\varepsilon^{\mu_1\mu_2\cdots \mu_n}\varepsilon_{\alpha_1\alpha_2\cdots \alpha_n}$. Throughout this work, we use the flat space metric $\eta_{\mu\nu}$ to raise and lower indices in quartic Lagrangians. We also find that in $D=4$, three of the spin two structures evaluate to zero and the only parity even CRG allowed Lagrangian is given by the zero derivative Lagrangian $\tilde{\mathcal{L}}_{\partial^0}^{\mathfrak{h}}$. This reduction, in some sense, obvious from the topological nature of our Lagrangians which can be expressed as product of Levi-Civita tensors. Note that the CRG allowed classification done in this paper is of a different nature than \cite{Bonifacio:2018vzv}, which sought to classify four point contact term in $D=4$ in the Gross-Mende limit (large $s$ and large $t$). The authors of this paper do a complete classification of both parity violating and parity preserving four point as well as three point couplings in $D=4$.

UV consistency bounds in massive gravity theories have also been of interest in recent years \cite{Bonifacio:2016wcb, Cheung:2016yqr, Hinterbichler:2017qyt, Boulanger:2018dau, Bellazzini:2019bzh, deRham:2017xox,Bellazzini:2017fep, Bonifacio:2018vzv, deRham:2018qqo, Alberte:2019xfh, Wang:2020xlt, Bellazzini:2023nqj}\footnote{See also \cite{Hinterbichler:2017qyt, Bonifacio:2017nnt, Edelstein:2021jyu} for causality motivated approaches to higher derivative theories of gravity with massive spin two particles.}. It is interesting to ask if we the space of such massive gravity theories are consistent in the ``classical" sense by imposing CRG. In this regard, we explicitly compute the tree-level amplitudes (both exchange diagams and contact term contributions) in dRGT gravity theory \cite{deRham:2010ik,deRham:2010kj}  where the amplitudes are a function of two parameters of the theory $(c_3, d_5)$ \footnote{We thank Shiraz Minwalla for pointing out this possibility to us.}. We now impose CRG on these tree-level amplitudes in the energy range, 
$$m^2\ll s\ll M^2,$$
where $M\sim (m^3 m_P)^\frac{1}{4}$ is the cutoff of the effective theory, $m$ is the mass of the massive spin two particle and $m_P$ is the reduced Planck mass. We find that there exist polarisation choices for which the tree-level S-matrix generated by the dRGT effective Lagrangian grows like $O(s^3)$ at large $s$ and fixed $t$ unless 
\begin{equation}
c_3=\frac{1}{4}.
\end{equation}
 
In this analysis, we have assumed that loop corrections are suppressed by $M$ so that CRG is applicable (See \cite{Park:2010rp, Park:2010zw, deRham:2013qqa} for explicit loop computations in ghost-free massive gravity where the one loop is suppressed compared to tree-level). This is also the conclusion of a causality based analysis by \cite{Camanho:2016opx}. While these bounds are not as strong as the positivity bounds of \cite{Bellazzini:2023nqj}, we feel CRG is an important criteria to construct tree-level effective Lagrangians of massive gravity as we explain below.

Since our analysis involving just the contact terms demonstrates that there do exist amplitudes which obey CRG, it would be worthwhile to consider CRG as a guiding principle to construct massive theories of gravity. It would be interesting to see, at a more abstract level, whether one can construct CRG obeying amplitudes that might be generated by a two derivative theories of massive spin two particle with ghost-free degrees of freedom. This would entail classifying all three point couplings of massive spin two field and restricting to six derivative contact term like couplings that we have classified. Now, we can take an arbitrary linear combination of these set of exchange amplitudes constructed from the classified three point couplings with the ghost-free massive spin two propagator and the contact amplitudes generated by the four point couplings. We would then look for linear combinations which obeys CRG. It would be interesting to see if such solutions satisfy the full UV consistency bounds as well. In this direction, the authors of \cite{Kundu:2023cof} have considered a single massive spin two field coupled to massless spin two field in $D=4$ and have explicitly shown no such theory can be consistent with CRG.

Consider the same Lagrangians listed  in eqs \eqref{introspinonelag} and \eqref{introspintwolag} but now in $AdS$ and let us study the four point CFT correlator that they generate at large $N$. These correlators, in the Regge limit, obey Chaos bound \cite{Maldacena:2015iua} and are solutions to large $N$ crossing equation involving just double trace operators in the CFT spectrum \cite{ Heemskerk:2009pn}. This has important consequences in the context of Inversion formula of a CFT correlator \cite{Caron-Huot:2017vep, Simmons-Duffin:2017nub}. In \cite{Turiaci:2018dht}, it was proven that scalar CFT correlators in large $N$ CFT can be completely determined by the double discontinuity of the correlator up to a fixed number of $AdS$ scalar contact term like interactions. Modulo some caveats which we explain in detail in \S\ref{inversionAmb}, we interpret the Lagrangians in eqs \eqref{spin1crgalwdlag} and \eqref{basisspin2id} as similar ambiguities in conformal inversion formula for four point correlators of unconserved spinning operators.

This paper is organized in the following way. In  \S\ref{kinematics} we define the kinematics of our process and explicitly state the unconstrained data in four point scattering after imposing equations of motion. In  \S\ref{cobs} we perform a group theoretic counting for the number of primaries and quasi-invariants. In  \S\ref{ecosms1} and \S\ref{ecosms2} we explicitly construct primaries and quasi-invariants for spin one and two respectively. Finally in section \S\ref{CRGstruc} we classify CRG allowed primaries as well as quasi-invariants and their descendants. In section \S\ref{4Dsct} we comment on our CRG allowed structures in $D=4$. In \S\ref{CRGdRGT}, we explicitly compute tree-level amplitudes in dRGT massive gravity theory and show that for all possible choices of the parameters in the theory, the amplitudes violate CRG. In section \S\ref{inversionAmb} explain the connection with the inversion formula of a large $N$ CFT. 
\section{Kinematics}\label{kinematics}

In the  $2 \rightarrow 2$ scattering processes that we study, the incoming and outgoing particles are  of the same spin but in general of different masses. We also generalise to the case where the particles of same mass. For non-identical particles, we can study twenty four different scattering processes (since we can permute the incoming and outgoing particles in twenty four different ways). It is therefore, convenient to introduce some notation to denote various scattering processes we study. We introduce the following notation for the scattering process where massive spin one and spin two particles of mass $m_i$ and $m_j$ are incoming and $m_k$ and $m_l$ are outgoing\footnote{$(i,j,k,l)$ takes values from various permutations of $(1,2,3,4)$.}, 

\begin{eqnarray}\label{ijtoklscat1}
    T_{\mathfrak{v}_i(p_1)\mathfrak{v}_j(p_2) \rightarrow \mathfrak{v}_k(p_3)\mathfrak{v}_l(p_4)}, \qquad T_{\mathfrak{h}_i(p_1)\mathfrak{h}_j(p_2) \rightarrow \mathfrak{h}_k(p_3)\mathfrak{h}_l(p_4)},
\end{eqnarray}
where $\mathfrak v$ and  $\mathfrak h$  denote spin one and spin two particles respectively. This scattering process obeys the following on-shell conditions and momentum conservation for the spinning particles,
\be\label{momonsc}
\begin{split}
&(p_{1})^2=-m_i^2,~~ (p_{2})^2=-m_j^2, ~~(p_{3})^2=-m_k^2,~~(p_{4})^2=-m_l^2,~~\sum_n\,  p^\mu_{n}=0.    
\end{split}
\ee

The Mandelstam invariants for this process are given by 
\begin{eqnarray}\label{stu12to34}
   &&s=-(p_1+p_2)^2=-2 p_1\cdot p_2 + m_i^2+m_j^2, \qquad t=-(p_1+p_3)^2=-2 p_1\cdot p_3 + m_i^2+m_k^2, \nonumber\\
   &&u=-(p_2+p_3)^2=-2 p_2\cdot p_3 + m_j^2+m_k^2, \qquad s+t+u = \sum_{i} m^2_i.
\end{eqnarray}

The different processes correspond to permuting the particles (labelled by different masses) while we continue to label the incoming and outgoing momenta by $p_1, p_2$ and $p_3, p_4$ respectively, it is convenient to represent the different processes pictorially. For example, the two different processes involving scattering of spin one particles, $ T_{\mathfrak{v}_1(p_1)\mathfrak{v}_2(p_2) \rightarrow \mathfrak{v}_3(p_3)\mathfrak{v}_4(p_4)}$ and 
$ T_{\mathfrak{v}_1(p_1)\mathfrak{v}_3(p_2) \rightarrow \mathfrak{v}_2(p_3)\mathfrak{v}_4(p_4)}$ can be pictorially represented as in figure \ref{1234fig} (time runs from left to right). For identical particles, of course such a distinction is not there and there is only one scattering process represented by, 
$ T_{\mathfrak{v}(p_1)\mathfrak{v}(p_2) \rightarrow \mathfrak{v}(p_3)\mathfrak{v}(p_4)}$. We  review the relevant degrees of freedom for massive spinning particles. 

\subsection*{Massive spin one}
Massive spin one particles are parametrised by their polarisation vectors, which obey the following free equations of motion \cite{PhysRevD.9.898}
\begin{eqnarray}
    {\epsilon_a}_\mu(p_i) {p_i}^\mu =0,\qquad (p_i^2 -m_a^2) {\epsilon_a}_{\mu}(p_i) =0,   
\end{eqnarray}
where $\epsilon_a$ is the polarisation vector of the spin one field of mass $m_a$\footnote{In order to avoid confusion, we use Latin letters to denote the type of particle and the Greek letters to denote the Lorentz indices.}. In the $2\rightarrow 2$ scattering experiment where the scattering plane is 3-dimensional (since $\sum^4_{i=1} p_i^\mu=0$ due to momentum conservation), we can divide polarisation in terms of  a three dimensional vector  $\polp_a$ in the scattering three plane spanned by the three independent momenta and a $(D-3)$ transverse vector $\polo_a$, orthogonal to the scattering plane. 
\begin{equation}\label{spinonepol}
    \epsilon_i= \polp_i+\polo_i,\qquad \textrm{where,}\qquad \polp_i\cdot\polo_i=0. \\
\end{equation}
In this equation, $\polo_i$ transforms as a vector of $SO(D-3)$. This decomposition is motivated by the fact that since the four-particle scattering happens in the three plane spanned by the momentum vectors, the scattering amplitude enjoys $SO(D-3)$ residual symmetry. Stated differently, $SO(D-3)$ is a stabilizer group of four point scattering amplitude. 

We now provide an explicit parametrization of $\polp_i$ in terms of momenta. For non-identical particles, solutions for our $\epsilon_a$ are process dependent. Let us consider the process,

 \begin{eqnarray}\label{12to34scat1}
    T_{\mathfrak{v}_1(p_1)\mathfrak{v}_2(p_2) \rightarrow \mathfrak{v}_3(p_3)\mathfrak{v}_4(p_4)}
\end{eqnarray}

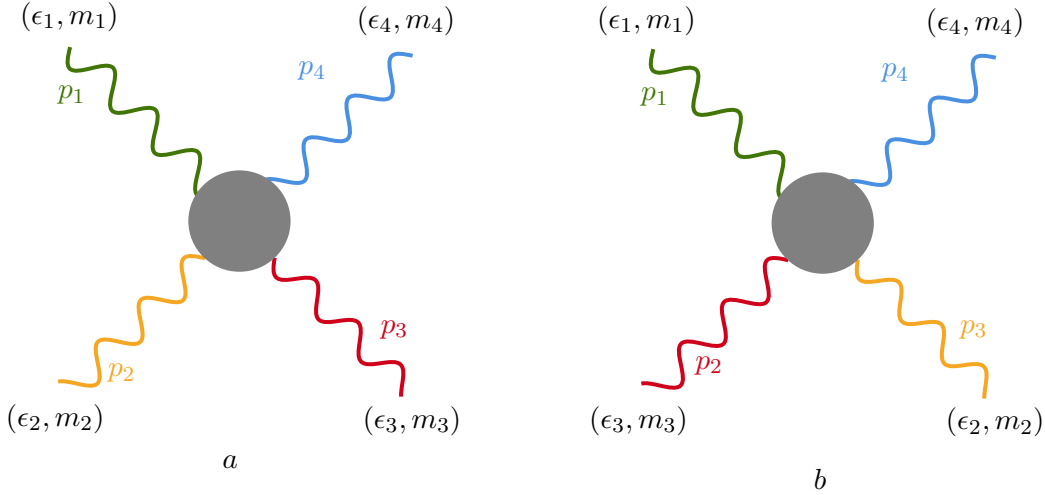
\begin{figure}
    \centering

\tikzset{every picture/.style={line width=0.75pt}} 

\begin{tikzpicture}[x=0.75pt,y=0.75pt,yscale=-1,xscale=1]

\draw  [color={rgb, 255:red, 65; green, 117; blue, 5 }  ,draw opacity=1 ][line width=1.5]  (110.5,113.5) .. controls (109.41,118.05) and (108.38,122.38) .. (110.3,124.3) .. controls (112.22,126.22) and (116.56,125.19) .. (121.11,124.11) .. controls (125.65,123.02) and (129.99,121.99) .. (131.91,123.91) .. controls (133.83,125.83) and (132.8,130.17) .. (131.71,134.71) .. controls (130.63,139.26) and (129.6,143.6) .. (131.52,145.52) .. controls (133.44,147.44) and (137.77,146.41) .. (142.32,145.32) .. controls (146.87,144.23) and (151.2,143.2) .. (153.12,145.12) .. controls (155.04,147.04) and (154.01,151.38) .. (152.93,155.93) .. controls (151.84,160.47) and (150.81,164.81) .. (152.73,166.73) .. controls (154.65,168.65) and (158.99,167.62) .. (163.53,166.53) .. controls (168.08,165.45) and (172.42,164.42) .. (174.34,166.34) .. controls (176.26,168.26) and (175.23,172.59) .. (174.14,177.14) .. controls (173.05,181.69) and (172.02,186.02) .. (173.94,187.94) .. controls (173.96,187.96) and (173.98,187.98) .. (174,188) ;
\draw  [color={rgb, 255:red, 208; green, 2; blue, 27 }  ,draw opacity=1 ][line width=1.5]  (212.55,219.45) .. controls (211.95,223.51) and (211.38,227.38) .. (213.3,229.3) .. controls (215.22,231.22) and (219.1,230.66) .. (223.15,230.06) .. controls (227.21,229.46) and (231.09,228.89) .. (233.01,230.81) .. controls (234.93,232.73) and (234.36,236.61) .. (233.76,240.67) .. controls (233.16,244.72) and (232.6,248.6) .. (234.52,250.52) .. controls (236.44,252.44) and (240.31,251.87) .. (244.37,251.27) .. controls (248.43,250.67) and (252.3,250.11) .. (254.22,252.03) .. controls (256.14,253.95) and (255.57,257.82) .. (254.97,261.88) .. controls (254.37,265.94) and (253.81,269.81) .. (255.73,271.73) .. controls (257.65,273.65) and (261.52,273.09) .. (265.58,272.48) .. controls (269.64,271.88) and (273.51,271.32) .. (275.43,273.24) .. controls (277.35,275.16) and (276.79,279.03) .. (276.19,283.09) .. controls (275.86,285.32) and (275.54,287.5) .. (275.64,289.32) ;
\draw  [color={rgb, 255:red, 74; green, 144; blue, 226 }  ,draw opacity=1 ][line width=1.5]  (281.38,117.84) .. controls (277.07,116.93) and (272.95,116.07) .. (271.01,117.97) .. controls (269.07,119.87) and (269.85,124) .. (270.67,128.34) .. controls (271.5,132.68) and (272.28,136.81) .. (270.34,138.71) .. controls (268.4,140.61) and (264.28,139.75) .. (259.96,138.84) .. controls (255.64,137.93) and (251.52,137.07) .. (249.58,138.97) .. controls (247.65,140.87) and (248.42,145) .. (249.25,149.34) .. controls (250.08,153.68) and (250.85,157.81) .. (248.92,159.71) .. controls (246.98,161.61) and (242.86,160.75) .. (238.54,159.84) .. controls (234.22,158.93) and (230.1,158.07) .. (228.16,159.97) .. controls (226.22,161.87) and (227,166) .. (227.83,170.34) .. controls (228.65,174.68) and (229.43,178.81) .. (227.49,180.71) .. controls (225.55,182.61) and (221.43,181.75) .. (217.11,180.84) .. controls (212.96,179.96) and (208.99,179.13) .. (206.97,180.77) ;
\draw  [color={rgb, 255:red, 245; green, 166; blue, 35 }  ,draw opacity=1 ][line width=1.5]  (177.81,219.71) .. controls (173.41,218.71) and (169.2,217.76) .. (167.27,219.66) .. controls (165.33,221.56) and (166.19,225.78) .. (167.1,230.21) .. controls (168.01,234.63) and (168.87,238.85) .. (166.93,240.75) .. controls (164.99,242.65) and (160.79,241.71) .. (156.39,240.71) .. controls (151.98,239.71) and (147.78,238.76) .. (145.84,240.66) .. controls (143.9,242.56) and (144.76,246.78) .. (145.68,251.21) .. controls (146.59,255.63) and (147.45,259.85) .. (145.51,261.75) .. controls (143.57,263.65) and (139.37,262.71) .. (134.96,261.71) .. controls (130.56,260.71) and (126.36,259.77) .. (124.42,261.67) .. controls (122.48,263.57) and (123.34,267.79) .. (124.25,272.21) .. controls (125.16,276.63) and (126.02,280.85) .. (124.09,282.75) .. controls (122.15,284.65) and (117.95,283.71) .. (113.54,282.71) .. controls (110.07,281.92) and (106.73,281.17) .. (104.52,281.81) ;
\draw  [color={rgb, 255:red, 128; green, 128; blue, 128 }  ,draw opacity=1 ][fill={rgb, 255:red, 128; green, 128; blue, 128 }  ,fill opacity=1 ] (170,201) .. controls (170,187.19) and (181.19,176) .. (195,176) .. controls (208.81,176) and (220,187.19) .. (220,201) .. controls (220,214.81) and (208.81,226) .. (195,226) .. controls (181.19,226) and (170,214.81) .. (170,201) -- cycle ;
\draw  [color={rgb, 255:red, 65; green, 117; blue, 5 }  ,draw opacity=1 ][line width=1.5]  (401.5,114.5) .. controls (400.41,119.05) and (399.38,123.38) .. (401.3,125.3) .. controls (403.22,127.22) and (407.56,126.19) .. (412.11,125.11) .. controls (416.65,124.02) and (420.99,122.99) .. (422.91,124.91) .. controls (424.83,126.83) and (423.8,131.17) .. (422.71,135.71) .. controls (421.63,140.26) and (420.6,144.6) .. (422.52,146.52) .. controls (424.44,148.44) and (428.77,147.41) .. (433.32,146.32) .. controls (437.87,145.23) and (442.2,144.2) .. (444.12,146.12) .. controls (446.04,148.04) and (445.01,152.38) .. (443.93,156.93) .. controls (442.84,161.47) and (441.81,165.81) .. (443.73,167.73) .. controls (445.65,169.65) and (449.99,168.62) .. (454.53,167.53) .. controls (459.08,166.45) and (463.42,165.42) .. (465.34,167.34) .. controls (467.26,169.26) and (466.23,173.59) .. (465.14,178.14) .. controls (464.05,182.69) and (463.02,187.02) .. (464.94,188.94) .. controls (464.96,188.96) and (464.98,188.98) .. (465,189) ;
\draw  [color={rgb, 255:red, 245; green, 166; blue, 35 }  ,draw opacity=1 ][line width=1.5]  (503.55,220.45) .. controls (502.95,224.51) and (502.38,228.38) .. (504.3,230.3) .. controls (506.22,232.22) and (510.1,231.66) .. (514.15,231.06) .. controls (518.21,230.46) and (522.09,229.89) .. (524.01,231.81) .. controls (525.93,233.73) and (525.36,237.61) .. (524.76,241.67) .. controls (524.16,245.72) and (523.6,249.6) .. (525.52,251.52) .. controls (527.44,253.44) and (531.31,252.87) .. (535.37,252.27) .. controls (539.43,251.67) and (543.3,251.11) .. (545.22,253.03) .. controls (547.14,254.95) and (546.57,258.82) .. (545.97,262.88) .. controls (545.37,266.94) and (544.81,270.81) .. (546.73,272.73) .. controls (548.65,274.65) and (552.52,274.09) .. (556.58,273.48) .. controls (560.64,272.88) and (564.51,272.32) .. (566.43,274.24) .. controls (568.35,276.16) and (567.79,280.03) .. (567.19,284.09) .. controls (566.86,286.32) and (566.54,288.5) .. (566.64,290.32) ;
\draw  [color={rgb, 255:red, 74; green, 144; blue, 226 }  ,draw opacity=1 ][line width=1.5]  (572.38,118.84) .. controls (568.07,117.93) and (563.95,117.07) .. (562.01,118.97) .. controls (560.07,120.87) and (560.85,125) .. (561.67,129.34) .. controls (562.5,133.68) and (563.28,137.81) .. (561.34,139.71) .. controls (559.4,141.61) and (555.28,140.75) .. (550.96,139.84) .. controls (546.64,138.93) and (542.52,138.07) .. (540.58,139.97) .. controls (538.65,141.87) and (539.42,146) .. (540.25,150.34) .. controls (541.08,154.68) and (541.85,158.81) .. (539.92,160.71) .. controls (537.98,162.61) and (533.86,161.75) .. (529.54,160.84) .. controls (525.22,159.93) and (521.1,159.07) .. (519.16,160.97) .. controls (517.22,162.87) and (518,167) .. (518.83,171.34) .. controls (519.65,175.68) and (520.43,179.81) .. (518.49,181.71) .. controls (516.55,183.61) and (512.43,182.75) .. (508.11,181.84) .. controls (503.96,180.96) and (499.99,180.13) .. (497.97,181.77) ;
\draw  [color={rgb, 255:red, 208; green, 2; blue, 27 }  ,draw opacity=1 ][line width=1.5]  (468.81,220.71) .. controls (464.41,219.71) and (460.2,218.76) .. (458.27,220.66) .. controls (456.33,222.56) and (457.19,226.78) .. (458.1,231.21) .. controls (459.01,235.63) and (459.87,239.85) .. (457.93,241.75) .. controls (455.99,243.65) and (451.79,242.71) .. (447.39,241.71) .. controls (442.98,240.71) and (438.78,239.76) .. (436.84,241.66) .. controls (434.9,243.56) and (435.76,247.78) .. (436.68,252.21) .. controls (437.59,256.63) and (438.45,260.85) .. (436.51,262.75) .. controls (434.57,264.65) and (430.37,263.71) .. (425.96,262.71) .. controls (421.56,261.71) and (417.36,260.77) .. (415.42,262.67) .. controls (413.48,264.57) and (414.34,268.79) .. (415.25,273.21) .. controls (416.16,277.63) and (417.02,281.85) .. (415.09,283.75) .. controls (413.15,285.65) and (408.95,284.71) .. (404.54,283.71) .. controls (401.07,282.92) and (397.73,282.17) .. (395.52,282.81) ;
\draw  [color={rgb, 255:red, 128; green, 128; blue, 128 }  ,draw opacity=1 ][fill={rgb, 255:red, 128; green, 128; blue, 128 }  ,fill opacity=1 ] (461,202) .. controls (461,188.19) and (472.19,177) .. (486,177) .. controls (499.81,177) and (511,188.19) .. (511,202) .. controls (511,215.81) and (499.81,227) .. (486,227) .. controls (472.19,227) and (461,215.81) .. (461,202) -- cycle ;

\draw (84,89.4) node [anchor=north west][inner sep=0.75pt]  [color={rgb, 255:red, 0; green, 0; blue, 0 }  ,opacity=1 ]  {$( \epsilon _{1} ,m_{1})$};
\draw (77,291.4) node [anchor=north west][inner sep=0.75pt]  [color={rgb, 255:red, 0; green, 0; blue, 0 }  ,opacity=1 ]  {$( \epsilon _{2} ,m_{2})$};
\draw (255,292.4) node [anchor=north west][inner sep=0.75pt]  [color={rgb, 255:red, 0; green, 0; blue, 0 }  ,opacity=1 ]  {$( \epsilon _{3} ,m_{3})$};
\draw (103,131.4) node [anchor=north west][inner sep=0.75pt]  [color={rgb, 255:red, 65; green, 117; blue, 5 }  ,opacity=1 ]  {$p_{1}$};
\draw (128,271.4) node [anchor=north west][inner sep=0.75pt]  [color={rgb, 255:red, 245; green, 166; blue, 35 }  ,opacity=1 ]  {$p_{2}$};
\draw (264,249.4) node [anchor=north west][inner sep=0.75pt]  [color={rgb, 255:red, 208; green, 2; blue, 27 }  ,opacity=1 ]  {$p_{3}$};
\draw (223,119.4) node [anchor=north west][inner sep=0.75pt]  [color={rgb, 255:red, 74; green, 144; blue, 226 }  ,opacity=1 ]  {$p_{4}$};
\draw (372,91.4) node [anchor=north west][inner sep=0.75pt]  [color={rgb, 255:red, 0; green, 0; blue, 0 }  ,opacity=1 ]  {$( \epsilon _{1} ,m_{1})$};
\draw (368,292.4) node [anchor=north west][inner sep=0.75pt]  [color={rgb, 255:red, 0; green, 0; blue, 0 }  ,opacity=1 ]  {$( \epsilon _{3} ,m_{3})$};
\draw (546,293.4) node [anchor=north west][inner sep=0.75pt]  [color={rgb, 255:red, 0; green, 0; blue, 0 }  ,opacity=1 ]  {$( \epsilon _{2} ,m_{2})$};
\draw (536,92.4) node [anchor=north west][inner sep=0.75pt]  [color={rgb, 255:red, 0; green, 0; blue, 0 }  ,opacity=1 ]  {$( \epsilon _{4} ,m_{4})$};
\draw (394,132.4) node [anchor=north west][inner sep=0.75pt]  [color={rgb, 255:red, 65; green, 117; blue, 5 }  ,opacity=1 ]  {$p_{1}$};
\draw (553,249.4) node [anchor=north west][inner sep=0.75pt]  [color={rgb, 255:red, 245; green, 166; blue, 35 }  ,opacity=1 ]  {$p_{3}$};
\draw (421,267.4) node [anchor=north west][inner sep=0.75pt]  [color={rgb, 255:red, 208; green, 2; blue, 27 }  ,opacity=1 ]  {$p_{2}$};
\draw (514,120.4) node [anchor=north west][inner sep=0.75pt]  [color={rgb, 255:red, 74; green, 144; blue, 226 }  ,opacity=1 ]  {$p_{4}$};
\draw (185,317.4) node [anchor=north west][inner sep=0.75pt]    {$a$};
\draw (480,324.4) node [anchor=north west][inner sep=0.75pt]    {$b$};
\draw (252,91.4) node [anchor=north west][inner sep=0.75pt]  [color={rgb, 255:red, 0; green, 0; blue, 0 }  ,opacity=1 ]  {$( \epsilon _{4} ,m_{4})$};

\end{tikzpicture}

    \caption{Two different scattering processes  $T_{\mathfrak{v}_1(p_1)\mathfrak{v}_2(p_2) \rightarrow \mathfrak{v}_3(p_3)\mathfrak{v}_4(p_4)}$ and $T_{\mathfrak{v}_1(p_1)\mathfrak{v}_3(p_2) \rightarrow \mathfrak{v}_2(p_3)\mathfrak{v}_4(p_4)}$ described by figure a and b respectively. The $\epsilon$ denote the polarisation tensor for the massive spin one particle.}
    \label{1234fig}
\end{figure}

This indicates the scattering process in which we have incoming spin one particles of mass $m_1$ and $m_2$ and the outgoing particles are of masses $m_3$ and $m_4$. The on-shell conditions (and momentum conservation) for this scattering process are,
\begin{eqnarray}
    \epsilon_i(p_i)\cdot p_i =0,\qquad p_i^2=-m_i^2,\qquad 
\end{eqnarray}
This process can be represented pictorially in subfigure a of figure \ref{1234fig}. In general, $\polp_1$ can be a function of all the three independent momenta, 
\begin{equation}
    \polp_1= a~\frac{p_1}{f(m_i,s,t,u)} +b~\frac{p_2}{g(m_i,s,t,u)} +c~\frac{p_4}{h(m_i,s,t,u)},
\end{equation}
for complex numbers $a,b$ and $c$ and functions $f,g$ and $h$ of the same dimensions as momenta for dimensional consistency. Recall, if the particle were massless, the term corresponding to $a$ would have been gauge redundant but since our external states are massive,  this corresponds to a physical polarisation, usually termed as the longitudinal polarisation. The space spanned by $p_2$ and $p_4$ is termed as transverse polarisation. Moreover, not all the three numbers are linearly independent and we have only two independent in-plane polarization vectors because of the equation of motion, $\polp_a \cdot p_i=0$ (since by construction, $\polo_a \cdot p_i=0$). We find that the most general solution of $\polp_1$, obtained in this manner, to be parametrized by two complex numbers. Denoting the transverse and longitudinal polarisations by $\epsilon_1^{T}$ and $\epsilon_1^{L}$ respectively, an explicit parametrization is given by,   
\begin{equation}\label{polpparaspinone}
    \begin{split}
         \polp_1&=\alpha_1 \epsilon_1^{L} + \beta_1 \epsilon_1^{T},\qquad \epsilon_1^{L}\cdot \epsilon_1^{T}=0,\qquad  |\epsilon_1^{L}|^2= |\epsilon_1^{T}|^2=1,\\
         \epsilon_1^{L} =& \mathcal{N}_1^{L}\left( \frac{p_1}{m_1} + \mathcal{C}_1 \frac{p_2}{m_1} +   \mathcal{C}_2 \frac{p_4}{m_1}\right),\qquad
         \epsilon_1^{T} = \mathcal{N}_1^{T}\left( \mathcal{T}_1 \frac{p_2}{m_1} -  \mathcal{T}_2 \frac{p_4}{m_1}\right),\\
             \end{split}
\end{equation}
where $\alpha_1$ and $\beta_1$ are arbitrary complex numbers and explicit expressions for  $\mathcal{N}_1^L, \mathcal{N}_1^T, \mathcal{C}_1, \mathcal{C}_2, \mathcal{T}_1$ and $ \mathcal{T}_2$ are given in eq \eqref{Normalisation} of Appendix \ref{eepv}. The two modes parametrized by $\alpha_1$ and $\beta_1$ corresponding to the longitudinal and the transverse modes respectively. Hence, in conclusion, the polarisation tensor for massive spin one particle is parametrized by $(\alpha_1, \beta_1, \polo_1)$. The rest of the polarizations ${\polp_2}(p_2),~{\polp_3}(p_3)$, and ${\polp_4}(p_4)$ can be obtained by just applying the double transposition elements $\Z_2 \otimes \Z_2 \in \left(P_{12}P_{34},P_{13}P_{24},P_{14}P_{23}\right)$  of the permutation group $S_4$ on the ${\polp_1}(p_1)$ polarisation vector, where $P_{ij}$ implies swap of the particle indices $i$ and $j$.\footnote{See appendix \ref{s3} for a brief review of the representation theory of the discrete group $S_4$ that we use in this paper.} If we denote the unit vectors in the $D-3$ dimensional orthogonal space to the scattering plane as $\hat{e}^a_{\mu}$, the most general massive spin one polarisation vector therefore is given by 
\begin{eqnarray}\label{eadef}
    &&{\epsilon}_\mu= \frac{1}{\sqrt{2}}\left(\alpha {\epsilon}_{\mu}^{L} + \beta 
    {\epsilon}_{\mu}^{T}\right) + {\polo}_\mu,\nonumber\\
    &&\sum\limits_{a=1}^{D-3}{e}^a_\mu={\polo}_\mu,\qquad e^a_\mu=e_a\hat{e}^a_\mu. \nonumber\\
\end{eqnarray}
Here $e_a$ are arbitrary numbers and we have suppressed the mass label for convenience. In Regge limit, (large $s$, fixed $t$) the leading $s$ behaviour of the $\polp_1$ is given by,

\begin{equation}
    \begin{split}
         {\polp_1}_\mu =&  \alpha_1 \left( \frac{{p_1}_\mu}{m_1} 
\right) + \beta_1  \left(\frac{{p_2}_\mu+{p_4}_\mu}{\sqrt{-t}}\right) 
         + \mathcal{O}\left(\frac{1}{s}\right) 
        \end{split}
\end{equation}

This is consistent with the fact that we expect the longitudinal mode to grow with energy while the transverse mode remains constant.

\subsection*{Massive spin two}
The spin two field is parametrized by a symmetric traceless polarisation tensor obeying the free equations of motion \cite{PhysRevD.9.898, Hinterbichler:2011tt}, 
\begin{eqnarray}\label{eomspin2}
    h^a_{\mu\nu}(p_i)p_i^\mu =0,\qquad (p_i^2 -m_a^2) h^a_{\mu\nu}(p_i) =0, \qquad {h^a}^\mu_\mu=0,   
\end{eqnarray}
where, similar to the spin one case, $h^a_{\mu\nu}$ is the polarisation vector of the spin two field of mass $m_a$. Suppressing the index associated with different masses, we can also express the polarisation tensor as,
\begin{equation}
    h_{\mu \nu}= \sum\limits_{i=1}^{\frac{D(D-1)}{2}} \varepsilon^{i}_{\mu\nu},\qquad {h}^\mu_\mu=0,   
\end{equation}

where $i$ runs over the basis elements of a symmetric $(D-1)\times (D-1)$ matrix and the index $\mu$ runs over $D$ spacetime dimensions. Without any loss of generality
, we choose to parameterize the massive spin two field polarisation as tensor product of two spin one polarisations  with the additional constraint of tracelessness. To be precise, we define the following two vectors in plane of three-dimensional $\polp$. \begin{equation}\label{polpparaspintwo2}
    \epsilon_{\mu}^{+}= \alpha \epsilon_{\mu}^{L} \quad \epsilon_{\mu}^{-}= \beta \epsilon_{\mu}^{T}
\end{equation}

The polarisations are therefore given by, 
\begin{itemize}\label{polpparaspintwo}
    \item  Three in-plane components of polarisation tensors determined by,
    \begin{eqnarray}
    \varepsilon_{\mu \nu}^{++}&=&\epsilon^{+}_{\mu}\epsilon^{+}_{\nu}, \qquad \varepsilon_{\mu\nu}^{-+}=  \frac{1}{2}\left(\epsilon^{-}_{\mu}\epsilon^{+}_{\nu}+\epsilon^{-}_{\nu}\epsilon^{+}_{\mu}\right),\qquad \varepsilon_{\mu \nu}^{--}=\epsilon^{-}_{\mu}\epsilon^{-}_{\nu}.
\end{eqnarray}

\item  Off diagonal terms are $2(D-3)$ in number and are of the form 
(see eq \eqref{eadef} for definition of $e^a_\mu$)
\begin{eqnarray}
\varepsilon_{\mu \nu}^{a \pm}=  \frac{1}{2}\left(\epsilon^{\pm}_{\mu}e^{a}_{\nu}+\epsilon^{\pm}_{\nu}e^{a}_{\mu}\right).
\end{eqnarray}

\item  A symmetric tensor of dimensions $\frac{(D-3)(D-2)}{2}$ composed of direct product of $\polo$, 
\begin{eqnarray}
    \varepsilon^{ab}_{\mu \nu}= e^a_\mu e^b_\nu.
\end{eqnarray}
\end{itemize}
We also impose the constraint of tracelessness which implies 
\begin{equation}
\begin{split}
     h^\mu_\mu &=\sum\limits_{i=1}^{\frac{D(D-1)}{2}} {\epsilon^i}_\mu^\mu=\alpha^2 + \beta^2+ \sum_{a=1}^{D-3}{e_a}^2=0.
\end{split}
    \end{equation}
In other words we can choose $h_{\mu \nu}= \epsilon_\mu \epsilon_\nu$ where $\epsilon_\mu$ is the spin one polarisation with the condition of tracelessness. In this expression we have used the following orthogonality relations 
\begin{equation}
\epsilon^{L}\cdot\epsilon^{T}=\hat{e}^a\cdot \hat{e}^b= \hat{e}^a\cdot \epsilon^{L/T}=0.
\end{equation}

The total degree of freedom is $\frac{(D+1)(D-2)}{2}$, which is indeed the degree of freedom for massive spin two particle as is evident from eq \eqref{eomspin2}. 
\begin{equation}
    N^{\mathfrak{h}}_{dof}= \frac{D(D+1)}{2}-D-1=\frac{(D+1)(D-2)}{2}.
\end{equation}

For the remainder of the paper, we choose to represent the scattering of four massive spin one and spin two particles using the following notation 

\begin{eqnarray}
    T^{{\mathfrak v}/{\mathfrak h}}_{\epsilon_i(p_1)\epsilon_j(p_2) \rightarrow \epsilon_k(p_3)\epsilon_l(p_4)}.
\end{eqnarray}

The generalisation of these solutions to the identical particles is straightforward by setting $m_i$ to $m$ in eq \eqref{polpparaspinone}.

\section{Group theoretic enumeration of amplitudes}\label{cobs}

Before we explicitly compute the analytic S-matrices, we first record our expectation from a group-theoretic analysis.  The enumeration of possible analytic structures contributing to tree-level $2 \rightarrow 2$ scattering of spinning particles can also be formulated as a group theory problem which has been addressed recently in great detail  for massless and massive particles in \cite{Henning:2015daa, Henning:2017fpj, Henning:2015alf, Chowdhury:2019kaq, Chowdhury:2020ddc, Chowdhury:2022xes}. 
The little group for massive spinning particles is $SO(D-1)$. Consider a particle whose polarisations transform in the irreducible representation $\rho$ of the little group. As explained in\S \ref{kinematics}, in a $2\rightarrow 2$ scattering experiment, the scattering plane is 3-dimensional and scattering amplitude has a residual $SO(D-3)$ symmetry. 
Therefore, the number of structures contributing to the scattering of four non-identical particles then becomes the number of $SO(D-3)$ singlets in $\rho^{\otimes 4}$ \cite{Chowdhury:2019kaq, Kravchuk:2016qvl}.
\begin{equation}\label{Nnonidentical}
    N_{non-id}=\rho^{\otimes 4}|_{SO(D-3)}.
\end{equation}

In order to compute this, it is convenient to restrict the $SO(D-1)$ representation $\rho_R$ to irreps of $SO(D-3)$.  

\begin{equation}
    \rho|_{SO(D-3)} = \oplus_g~ n_g {\tilde{\rho}_g},
\end{equation}
where $\tilde{\rho}_g$ represents irreducible representations of  $SO(D-3)$, $n_g$ represents their degeneracy and the sum is over all possible irreducible representations that appear in such a restriction.  The explicit structures are the $SO(D-3)$ invariant polynomials of $\tilde{\rho}^i_g$ with appropriate homogeneity where $i$ labels the particles. The restriction of a generic $SO(M)$ irreducible representation to irreps of $SO(M-1)$ is a mathematical problem whose solution is given by ``Branching rules", reviewed in great detail in the context of scattering amplitudes in  \cite{Chakraborty:2020rxf} and we will use their results extensively in this section.

If the particles are identical, the amplitudes enjoy an additional $S_4$ symmetry ( the symmetry group pertaining to permutations of four objects, see appendix \ref{s3}). Recall that $S_4$ group has an normal subgroup $\mathbb{Z}_2 \times\mathbb{Z}_2$, and $S_4$ can be written as the semi-direct product of $S_3$ and $\mathbb{Z}_2\times \mathbb{Z}_2$.

$$S_4 \cong S_3 \ltimes (\mathbb{Z}_2\times \mathbb{Z}_2).$$

We first impose the $\mathbb{Z}_2\times \mathbb{Z}_2$ symmetry on our polynomial of $\tilde{\rho}^i_g$. The space of such structures are called ``quasi-invariant" S-matrices \cite{Chowdhury:2019kaq}. The counting of such structures is given by , 

\begin{equation}\label{quasiinv}
     N_{Quasi-inv}=\rho^{\otimes 4}|_{\mathbb{Z}_2 \times \mathbb{Z}_2} = \rho^{\otimes 4} - 3S^2\rho \otimes \wedge^2 \rho,\qquad \rho\equiv\rho|_{SO(D-3)} = \oplus_g~ n_g {\tilde{\rho}_g},  
\end{equation}
where $S^2$ and $\wedge^2$ denote the symmetric and the anti-symmetric tensor product and the singlet condition for $SO(D-3)$ is implied. The quasi-invariant structures can further be organised into irreducible representations of $S_3$. As reviewed in Appendix \ref{s3}, there are three irreducible representations of $S_3$,  
\begin{equation}
    {\bf 1_S}={\tyng(3)},\qquad {\bf 2_M}={\tyng(2,1)},\qquad {\bf 1_A}={\tyng(1,1,1)}.
\end{equation}

These constitute the one dimensional symmetric and anti-symmetric representations of $S_3$ and the two dimensional mixed symmetric representation. We now enumerate amplitudes for the cases of  scattering of massive spin one and massive spin two particles. Note that in this work, we will mostly consider the space-time dimension $D\geq 8$. The group theory and the subsequent enumeration of structures are only valid in this context. For lower dimensions, we can have reduction of the structures we classify in this work as well as appearance of additional parity violating structures in specific dimensions (Additional Lorentz invariant structures  can be formed by contraction with space time Levi-Civita tensor in lower dimensions).

\subsection{Massive spin one}\label{spin1count}
The polarization parametrizing a massive spin one particle  transforms under vector representation of $SO(D-1)$ in $D$ dimensional space-time. A  $SO(D-1)$ vector  transforms as a direct sum of a $SO(D-3)$ vector and two $SO(D-3)$ scalars.
\begin{equation}\label{rho1}
    \bf \rho_{\mathfrak{v}} =  {\tyng(1)} \oplus  2~{\bullet}.
\end{equation}

This is consistent with our analysis regarding in-plane and orthogonal decomposition of the massive spin one polarisation vector around eq \eqref{eadef}. We identify the two scalars as $\alpha, \beta$ while the vector is identified with $\polo$.   

\subsubsection{Scattering amplitudes for non-identical particles}
In order to  evaluate eq \eqref{Nnonidentical} it is convenient to  enumerate the direct product $\rho_{\mathfrak{v}} \otimes \rho_{\mathfrak{v}}$. 
\begin{eqnarray}\label{rho4singlets}
 \rho_{\mathfrak{v}} \otimes \rho_{\mathfrak{v}} = \oplus_R\, n_R\, \rho_R = \oplus_R\, (n^S_R + n^A_R)\, \rho_R,  
\end{eqnarray}
where $n_R$ deontes the degeneracy of the $SO(D-3)$ irreducible representation $\rho_R$ that appears in the tensor product. The degeneracy $n_R$ that appears for each representation $R$ in eq \eqref{rho4singlets}, is a sum of the degeneracies of the representation $R$ appearing in the symmetric and the anti-symmetric product of the two representations $\rho_{\mathfrak{v}}$. The number of $SO(D-3)$ singlets in $\otimes^4 \rho_{\mathfrak{v}}$ is then easily obtained,

\begin{eqnarray}
    N^{\mathfrak{v}}_{non-id}= \sum_R~(n^S_R+ n^A_R)^2.
\end{eqnarray}

For vectors transforming in the little group of massive spin one, we group the product into irreducible representations arising out of symmetric and anti-symmetric product of the two representations in table \ref{spin1Z2z2}.
\begin{table}[h!]

\centering
\begin{tabular}{|c|c|c|}
    \hline
   Irreps & $S^2 \rho$ &   $\Lambda ^2 \rho$\\
   \hline
   ${\tyng(2)}$ &  1 & 0\\
   \hline
   ${\tyng(1,1)}$ &  0 & 1\\
   \hline
   $\bullet$ &  4 & 1\\
   \hline
   {\tyng(1)} & 2 & 2 \\
     \hline
\end{tabular}
\caption{ Product of two vector space of massive spin $1$ polarizations.}
\label{spin1Z2z2}
\end{table}

From eq \eqref{rho4singlets} we easily see the number of singlets is, 
\begin{equation}\label{rho4singlestsspin1}
N^{\mathfrak{v}}_{non-id}= 1^2 + 1^2 + (4+1)^2 + (2+2)^2= 43.    
\end{equation}
 We verify this counting in the later sections by explicitly constructing the independent S-matrices for the massive spin one case in \S\ref{ecosms1}.

\subsubsection{Scattering amplitudes for identical particles}

For identical particles, using eq \eqref{quasiinv}, we can count the number of S-matrices which are $\mathbb{Z}_2\times \mathbb{Z}_2$ invariants \cite{Chowdhury:2019kaq, Kravchuk:2016qvl},

\begin{equation}\label{rho4z2z2}
    \rho_{\mathfrak{v}}^{\otimes 4}|_{\mathbb{Z}_2 \times \mathbb{Z}_2} = \rho_{\mathfrak{v}}^{\otimes 4} - 3S^2\rho_{\mathfrak{v}} \otimes \Lambda^2 \rho_{\mathfrak{v}},
\end{equation} 

The number of tensor structures in the symmetric and antisymmetric tensor product of $\rho$ is given in the table \ref{spin1Z2z2}. The number of $\mathbb{Z}_2 \times \mathbb{Z}_2$ invariant spin one S-matrices is therefore 
\begin{eqnarray}
N^{\mathfrak{v}}_{Quasi-inv}= (43 - 3(4\times 1) - 3(2\times2)) = 19,   
\end{eqnarray}
where we have used eq \eqref{rho4singlestsspin1}. 
\subsection{Massive spin two}

In $D$ dimensional space-time, a massive spin two particle transforms in a symmetric traceless tensor representation of the little group $SO(D-1)$. The symmetric traceless $SO(D-1)$ tensor transforms as 2 vectors, 3 scalars, and a symmetric traceless tensor representation of $SO(D-3)$. 

\begin{equation}
    \bf \rho_{\mathfrak{h}} = {\tyng(2)} \oplus 2\, {\tyng(1)} \oplus \
3 {\bullet} .
\end{equation}
This is consistent with our analysis around eq \eqref{polpparaspintwo}. We identify the three scalars as parametrizing the tensors $\varepsilon_{\mu\nu}^{++}, \varepsilon_{\mu\nu}^{-+}, \varepsilon_{\mu\nu}^{--}$, the two vectors parametrize $\varepsilon_{\mu\nu}^{+a}, \varepsilon_{\mu\nu}^{-a}$, while the symmetric traceless tensor parametrizes the components of $\varepsilon_{\mu\nu}^{ab}$ without the trace.

As a consistency check we show that the degrees of freedom encoded in this decomposition is as expected for a $SO(D-1)$ symmetric traceless tensor. Denoting by $d[R]$, the dimensions of the representation $R$ of $SO(D-3)$,  
\begin{eqnarray}
d[{\tyng(2)}]+2d[{\tyng(1)}]+3 d[{\bullet}]&=& \frac{(D-2)(D+1)}{2},
\end{eqnarray}
which is the expected degree of freedom for a symmetric traceless $SO(D-1)$ tensor.

\subsubsection{Scattering amplitudes for non-identical particles}
As before, we find the Clebsch-Gordon decomposition of the tensor product $\rho ^{\otimes 2}$. There are now nine irreps appearing the tensor product. The organisation of their degeneracies in the symmetric and the anti-symmetric sector is given in table \ref{spin2z2z2}.
\begin{table}[h]

\centering
\begin{tabular}{|c|c|c|}
    \hline
   Irreps & $S^2 \rho$ &   $\Lambda ^2 \rho$\\
   \hline
   ${\tyng(4)}$ & 1 & 0 \\
    \hline 
    ${\tyng(3,1)}$ & 0 & 1 \\
    \hline 
    ${\tyng(2,2)}$ & 1 & 0\\
    \hline 
    ${\tyng(3)}$ &  2 & 2\\
 \hline  
   ${\tyng(2,1)}$ & 2 & 2 \\
   \hline 
    ${\tyng(1)}$ & 8 & 8 \\
     \hline 
       ${\tyng(2)}$ & 7 & 4 \\
    \hline 
     ${\tyng(1,1)}$ & 1 & 4 \\
     \hline
      ${\bullet}$ & 10 & 4 \\
    \hline 
\end{tabular}
\caption{ Product of two symmetric traceless representation of massive spin $2$ polarizations.}
\label{spin2z2z2}
\end{table}
Number of structures contributing to non-identical scattering is given by, 

   \begin{equation}\label{rho4singlestsspin2}
N^{\mathfrak{h}}_{non-id} = 633.
\end{equation}

\subsubsection{Scattering amplitudes for identical particles}

For identical particles, we can enumerate the quasi-invariant structures using eq \eqref{quasiinv}, 
\begin{eqnarray}
N^{\mathfrak{h}}_{Quasi-inv}= 201.  \nonumber\\  
\end{eqnarray}

We explicitly construct these structures in \S\ref{ecosms2}.

\section{Explicit construction of the S-matrices: spin one}\label{ecosms1}

In this section we explicitly construct the contact tree-level amplitudes that contribute to scattering process involving four non-identical spin one particles and also impose $S_4$ symmetry for the situation when the particles are identical. Massive spinning particles do not have any gauge symmetry and since, at first, we are considering non-identical particles, we don't need to consider the symmetry of  S-matrices under the permutation group $S_4$. The allowed local S-matrices are therefore easy to construct. The massive spin one fields admit plane wave solutions, parametrized by the polarisation tensor $\epsilon_a^\mu(p_i)$, which obeys the following condition. 
\begin{eqnarray}
    \epsilon_a({p_i})\cdot p_i=0,\qquad {p_i}^2= -m_a^2
\end{eqnarray}
where $a$ denotes a particle of mass $m_a$.

Let us consider the scattering process in which particles labelled by the polarisation vectors $\epsilon_1^\mu$ and  $\epsilon_2^\mu$ are the incoming particles and  particles corresponding to $\epsilon_3^\mu$ and  $\epsilon_4^\mu$ are the outgoing particles. In terms of the notation introduced in eq \eqref{ijtoklscat1}, 

\begin{eqnarray}\label{12to34scat}
    T^{\mathfrak{v}}_{\epsilon_1(p_1)\epsilon_2(p_2) \rightarrow \epsilon_3(p_3)\epsilon_4(p_4)}
\end{eqnarray}
The on-shell conditions (and momentum conservation) for this scattering process are,
\begin{eqnarray}\label{12to34onshell}
    \epsilon_i(p_i)\cdot p_i =0,\qquad p_i^2=-m_i^2,\qquad 
\end{eqnarray}
Since the particle mass label and the momenta index label coincide, we also represent this as 
\begin{equation}
    T^{\mathfrak{v}}_{12 \rightarrow 34}
\end{equation}

For the scattering process in eq \eqref{12to34scat}, we can construct all independent local S-matrices by using independent scalar data which we determine as follows. Using equations of motion, eq \eqref{12to34onshell} and momentum conservation, the linearly independent data constitutes,

\begin{equation}\label{VarDef}
\begin{split}
A_{12} = \epsilon_1.p_2, ~~~~ & A_{14} = \epsilon_1.p_4, ~~~~  A_{21} = \epsilon_2.p_1,\\
A_{23} = \epsilon_2.p_3, ~~~~ & A_{32} = \epsilon_3.p_2, ~~~~  A_{34} = \epsilon_3.p_4,\\
A_{41} = \epsilon_4.p_1, ~~~~ & A_{43} = \epsilon_4.p_3, ~~~~ b_{ij} = \epsilon_i.\epsilon_j,\\
\end{split}
\end{equation}

where, $\epsilon_i\equiv \epsilon_i(p_i)$, and $p_i$ are the polarisation and the momentum of the $i^{th}$ spin one particle. The reader can easily check that other $A_{ij}= \epsilon_i. p_j$ can be reduced to the set in eq \eqref{VarDef}
using eq \eqref{12to34onshell} and momentum conservation. For notational convenience, we name $\epsilon_i. p_j$ independent structures for each $i$ as follows,
\begin{equation}\label{xidef}
    \begin{split}
        &\epsilon_1 . p_2 = \alpha_1, \quad \epsilon_1 . p_4 = \beta_1, \quad \epsilon_3 . p_2 = \beta_3, \quad \epsilon_3 . p_4 = \alpha_3,\\
        &\epsilon_2 . p_1 = \alpha_2, \quad \epsilon_2 . p_3 = \beta_2,\quad \epsilon_4 . p_1 = \beta_4, \quad \epsilon_4 . p_3 = \alpha_4.\\
    \end{split}
\end{equation}
 For the rest of the paper, we will denote $A_{i,j}$ by $\xi_i$, which can take two values corresponding to the two different values of $j$ index- $\alpha_i$ and $\beta_i$. Thus our S-matrices are local polynomials of $A_{ij}, \alpha_i$ and $\beta_i$ with appropriate homogeneity and graded by derivatives. We start at the lowest derivative order i.e derivative order zero and move to higher derivative orders, looking for linearly independent polynomials of $A_{ij}, b_{ij}$ with the correct homogeneity at each derivative order. Such polynomials without any symmetry, where the derivative order is counted through the number of $A_{ij}$s in the structure, i.e we don't have explicit Mandelstam variables in the structures, are called primaries.
 
 \subsection{S-matrices as module and descendants}\label{samd}
 
 The process outlined in the previous subsection is finite and the polynomials of $A_{i,j}$ and $b_{ij}$, classified in this manner, are termed generators of `primaries. The space of such polynomials can further be multiplied by local polynomials of Mandelstam invariants, which are termed ``descendants". Analytic S-matrices form a module rather rather than a vector space over the ring of polynomials generated by $s, t$ and $m$. Note that pole exchanges imply multiplication with non-analytic polynomials in momenta and hence are not descendants of the primaries.   
 
 For identical particles we have to impose additional $S_4$ symmetry on the space of primaries and there is a nice structure which emerges \cite{Henning:2017fpj, Chowdhury:2019kaq}. We impose $S_4$ invariance on the space of primaries in two steps. The first step involves imposing $\Z_2 \times \Z_2$ invariance to obtain quasi-invariants which generate the ``local module" for identical scattering. The local module can further be organised into irreducible representations of $S_3$. The symmetric and the anti-symmetric module are one dimensional and are denoted by $e^{\partial^l}_{\bf 1_S}$ and $e^{\partial^l}_{\bf 1_A}$ respectively where the superscript $\partial^l$ denotes the derivative order of the module. The mixed symmetric modules, on the other hand, are two dimensional and are represented as a triplet $\{e^{\partial^l, (1)}_{\bf 2_M}, e^{\partial^l, (2)}_{\bf 2_M}, e^{\partial^l, (3)}_{\bf 2_M}\}$ with the following condition 
between its orbit elements, 
$$e^{\partial^l, (1)}_{\bf 2_M}+ e^{\partial^l, (2)}_{\bf 2_M}+ e^{\partial^l, (3)}_{\bf 2_M}=0.$$

Following the conventions of \cite{Chowdhury:2019kaq}, we canonically choose the orbit element $e^{\partial^l, (1)}_{\bf 2_M}$ to be symmetric under $P_{12}$, $e^{\partial^l, (2)}_{\bf 2_M}$ to be symmetric under $P_{23}$ and $e^{\partial^l, (3)}_{\bf 2_M}$ to be symmetric under $P_{13}$. An efficient way of generating S-matrices from the quasi-invariant modules was outlined in \cite{Chowdhury:2019kaq}. Given the set of polynomials of $A_{i,j}$ and $b_{i,j}$ with no symmetry, we impose $S_4$ symmetry in two steps. We first impose $\Z_2 \otimes \Z_2$ symmetry using the projector eq \eqref{z2z2proj} in Appendix \ref{s3} 
\begin{equation}
    \Pi_0 = \frac{1}{4}\left(1 + P_{12}P_{34}+P_{13}P_{24}+P_{14}P_{23}\right),
\end{equation}
and find the set of linearly independent quasi-invariants. Given the set of quasi-invariant S-matrices $\{T^{QI}_i\}$ we find the linearly independent set of tensor structures in the set $\{\Pi_R \left(T^{QI}_i\right)\}$, where the projectors have been worked out in eq \eqref{s3proj},

\begin{equation}
    \begin{split}
        &\Pi_{\mathbf{1_S}} = \frac{1}{6}\left(1 + P_{12} + P_{23} + P_{13} + P_{132} + P_{123}\right),\\&
        \Pi_{\mathbf{1_A}} = \frac{1}{6}\left(1 - P_{12} - P_{23} - P_{13} + P_{132} + P_{123}\right),\\&
        \Pi^{(1)}_{\mathbf{2_M}} = \frac{1 + P_{12}}{2} - \Pi_{\mathbf 1_S},\\&
        \Pi^{(2)}_{\mathbf{2_M}} = \frac{P_{23}+P_{132}}{2} - \Pi_{\mathbf 1_S}.
    \end{split}
\end{equation} 

This  enumerates the number of linearly independent quasi-invariant S-matrices transforming in the irreducible representation $R$ and their orbits. For tensor structures transforming in $2_M$ the other element in the orbit is generated by the action of the  $\Pi^{(2)}_{\mathbf{2_M}}$ or $\Pi^{(3)}_{\mathbf{2_M}}$ (see eq \eqref{s3proj}).
These quasi invariants form the generators of the module in the sense that all the higher derivative structures are obtained from these generators by multiplication of appropriate polynomials of Mandelstam invariants. To be precise, given a quasi-invariant module of derivative  order $l$, which transforms in the irreducible representation ${\bf R}$ of $S_3$, the $\text{n}^{\text{th}}$ derivative descendant module are generated by multiplying a polynomial of derivative order $n-l$ transforming in the same irrep ${\bf R}$. The S-matrix generated by a generator module at any derivative order are therefore obtained from these descendant quasi-invariant modules by projection onto $S_3$ singlets. The polynomial ring, which generates the descendants, corresponding to each irreducible sector has been classified for the massless case \cite{Chowdhury:2019kaq} and the same classification carries over to the massive case with a minor modification since we have an additional scale in the problem due to the mass ``m" of the identical particles. Besides this minor modification, the most general descendant S-matrices for each irreducible sector can be generated as follows,

\begin{equation}\label{mostgendescl1}
\begin{split}
&\mathcal{S}^{n}_{{\bf 1_S}}= m^a(s^2+t^2+u^2)^b (s t u)^c e^{\partial^l}_{{\bf 1_S}},\\
&\mathcal{S}^{n+2}_{{\bf 2_M},1}= m^a(s^2+t^2+u^2)^b (s t u)^c \left((2s-t-u)e^{\partial^l,(1)}_{{\bf 2_M}}+(2t-s-u)e^{\partial^l,(2)}_{{\bf 2_M}}+(2u-t-s)e^{\partial^l,(3)}_{{\bf 2_M}}\right),\\
&\mathcal{S}^{n+4}_{{\bf 2_M},2}= m^a(s^2+t^2+u^2)^b (s t u)^c \left((2s^2-t^2-u^2)e^{\partial^l,(1)}_{{\bf 2_M}}+(2t^2-s^2-u^2)e^{\partial^l,(2)}_{{\bf 2_M}}+(2u^2-t^2-s^2)e^{\partial^l,(3)}_{{\bf 2_M}}\right),\\
&\mathcal{S}^{n+6}_{{\bf 1_A}}= m^a(s^2+t^2+u^2)^b (s t u)^c \left(s^2t-t^2s+t^2u-u^2t+u^2s-s^2u\right)e^{\partial^{l}}_{{\bf 1_A}},\\
\end{split} 
\end{equation}
where $n=a+4b+6c+l$. In particular note that the S-matrices generated from a module that transforms in ${\bf 2_M}$ or ${\bf 1_A}$ must be a higher derivative descendant. 

The amplitudes classified in this manner are in one-to-one correspondence with the equivalence class of local Lagrangians \cite{Chowdhury:2019kaq}. Two Lagrangians differing by equations of motion and total derivatives, give rise to the same scattering amplitude. The generators of the local module correspond to the linearly independent Lagrangians at lowest order in derivatives. Descendants of the local modules are generated by putting contracted derivatives on different fields in the Lagrangians which span the basis of local modules. These contracted derivatives in the momentum space correspond to polynomials of Mandelstam invariants multiplying the generators of local module. In this work, we will not attempt to list the explicit local Lagrangians or their descendants that generate these amplitudes  except in special cases. In this subsection and the next we provide the non-identical and identical module construction in terms of possible polynomial structure for each derivative order, relegating the explicit structure to the \texttt{Mathematica} files \textit{spinone\_regge.nb} (and \textit{spintwo\_regge.nb} for spin two)

\subsection{ Non-identical particles}

We classify the amplitudes corresponding to the non-identical scattering corresponding to the process, 
\begin{eqnarray}
    T^{\mathfrak{v}}_{\epsilon_1(p_1)\epsilon_2(p_2) \rightarrow \epsilon_3(p_3)\epsilon_4(p_4)}
\end{eqnarray}
This construction is accomplished in two steps. In the first step, we saturate the homogeneity in polarisation by counting the number of structures consisting of $\xi$ and $b_{ij}$. In the second step, we count the degeneracy of each such structure keeping in mind that each $\xi_i$ can take two values $\{\alpha_i, \beta_i\}$.  

 \subsection*{Zero derivative  S-matrices:}
Structures at zero derivatives are just polynomials of $b_{ij}$. It is not hard to convince oneself that homogeneity requirement ensures that only one kind of structure is possible 
\begin{eqnarray}
    M^{\mathfrak{v}}_{\partial^0, 1}= b_{ij} b_{kl}, \qquad \forall~~ i,j,k,l= 1,\cdots 4, 
\end{eqnarray}
where the superscript denotes this is a structure corresponding to spin one scattering and the subscript denotes the derivative order of the structure and labels the degeneracy. Taking into account the symmetricity of $b_{ij}$, we can have only three possible structures
\begin{equation}\label{zroder}
    \begin{split}
        b_{12} b_{34},\quad
        b_{13} b_{24},\quad
        b_{14} b_{23}.
    \end{split}
\end{equation}

\subsection*{Two derivative  S-matrices:}
For the two derivative S-matrices, there can be one possible tensor structure, 

\begin{equation}\label{tspinone2der}
     M^{\mathfrak{v}}_{\partial^2, 1}= \xi_i \xi_j b_{kl}, 
\end{equation}
where the indices $i,j,k,l= 1,\cdots 4$ taking into account homogeneity of the polarisation tensors. There are ${4 \choose 2}$ ways of selecting the $i,j$ indices and each of the $\xi_i$s can take two values. This brings the total number of two derivative structures to $6 \times 2^2=24$. 

\subsection*{Four derivative  S-matrices:}

For the  four derivative S-matrices, again, there can be one possible tensor structure, 

\begin{equation}\label{tspinone4der}
     M^{\mathfrak{v}}_{\partial^4, 1}= \xi_i \xi_j \xi_k \xi_l, \qquad \forall~~ i,j,k,l= 1,\cdots 4, 
\end{equation}
taking into account the homogeneity of the polarisation tensors. Each of the $\xi_i$s can take two values, which  brings the total number of two derivative structures to $2^4=16$.

We have forty three massive spin one S-matrices, as expected from our group theoretic counting in eq \eqref{rho4singlestsspin1}.

\subsection{Identical particles}\label{irrepsspin1}

In this subsection, we specialise to the case when the four particles are identical and the amplitudes have an additional $S_4$ permutation symmetry. As explained in\S \ref{samd}, we impose $S_4$ invariance in two steps. We first impose $\mathbb{Z}_2 \times \mathbb{Z}_2$ symmetry and enumerate the quasi-invariant S-matrices. In the second step we classify the amplitudes into irreducible irreps of $S_3$. 
\subsection*{ Zero derivative quasi-invariant S-matrices}

Quasi-invariant S-matrices are invariant under the double transposition generated by\\ $(P_{12}P_{34}, P_{13}P_{34})$. We record the behaviour of our building blocks under the double transposition. For the building block $b_{ij}$, we have two possibilities depending on which $\Z_2$ cycle of $\Z_2 \times \Z_2$ it is invariant under. For example,   
\begin{equation}\label{z2z2bij}
    \begin{split}
           P_{12}P_{34}(b_{12})=b_{12},\qquad 
             P_{13}P_{24}(b_{12})=P_{14}P_{23}(b_{12})=b_{34},\\
         \end{split}
\end{equation}
where $\left( P_{12}P_{34},~ P_{13}P_{24},~ P_{14}P_{23}\right)$  are the elements of $\Z_2 \times \Z_2$ and $ i,j\neq k, l$. For each $b_{ij}$, therefore it remains invariant under one  element of $\Z_2 \times \Z_2$. Zero derivative structures in eq \eqref{zroder} are, hence, already $\Z_2 \times \Z_2$ invariant and they are three in number. The zero derivative quasi-invariant structures transform in a ${\bf 1_S} + {\bf 2_M}$ representation of $S_3$.
\begin{equation}
    3= {\bf 1_S}+ {\bf 2_M}.
\end{equation}
The generator of the quasi-invariant module transforming in ${\bf 1_S}$ and ${\bf 2_M}$ are respectively given by 
\begin{equation}
    \begin{split}
        e^{\partial^0}_{{\bf 1_S}}&= \frac{1}{3}(b_{1,2} b_{3,4}+b_{1,3}b_{2,4}+b_{1,4}b_{2,3})\\
    e^{\partial^0, (1)}_{{\bf 2_M}}&= \frac{1}{3}(2b_{1,2} b_{3,4}-b_{1,3}b_{2,4}-b_{1,4}b_{2,3})\\
    \end{split}
\end{equation}

\subsection*{ Two derivative quasi-invariant S-matrices}

 The other building block, $\xi_i$, was used to denote two possibilities $\xi_i\in (\alpha_i, \beta_i)$ in eq \eqref{VarDef}.  We explicitly see from eq \eqref{VarDef}, 
\begin{equation}\label{z2z2aibi}
    \begin{split}
        P_{ij}P_{kl}(\alpha_{i})= \alpha_j, \qquad  P_{ij}P_{kl}(\beta_{i})= \beta_j.
    \end{split}
\end{equation}
Recall the two derivative structures we enumerated in eq \eqref{tspinone2der}. 
\begin{equation}
     M^{\mathfrak{v}}_{\partial^2, 1}= \xi_i \xi_j b_{kl}, \qquad \forall~~ i,j,k,l= 1,\cdots 4. 
\end{equation}
Our analysis in eqs \eqref{z2z2bij} and \eqref{z2z2aibi} imply that the two derivative Quasi-invariant S-matrices must be of the form. 

\begin{equation}
     M^{\mathfrak{v}}_{\partial^2, 1}|_{\Z_2 \times \Z_2}= \left(\xi_i \xi_j b_{kl} + \xi_k \xi_l b_{ij}\right), \qquad \forall~~ i,j,k,l= 1,\cdots 4 
\end{equation}
This combination is invariant under all elements  $P_{\Z_2\times\Z_2} \in \left(P_{12}P_{34},~ P_{13}P_{24},~ P_{14}P_{23}\right),$ for all $i\neq j \neq k \neq l$.
We can choose the indices $i,j$ in ${4 \choose 2}/2=3$ ways because of the $\Z_2 \times \Z_2$ symmetry. Recall that all the $b_{ij}$s are invariant under one of the elements of $\Z_2 \times \Z_2$. Hence the choices for $\xi_i \xi_j$ that leaves this combination invariant under the $\Z_2 \times \Z_2$ element, which leaves $b_{kl}$ invariant, amounts to three possibilities (because of eq \eqref{z2z2aibi})
\begin{equation}\label{xisqposs}
    \alpha_i \alpha_j,\qquad \beta_i \beta_j, \qquad \left(\alpha_i \beta_j +\alpha_j \beta_i\right).
\end{equation}
Hence the total counting of quasi-invariant S-matrices at two derivative order is 9. The $S_3$ representation theory is given by,

\begin{equation}\label{twoders3spinone}
9=  2~{\bf 1_S} + {\bf 1_A} + 3~{\bf 2_M}.
\end{equation}

\subsection*{ Four derivative quasi-invariant S-matrices}

Similarly, quasi-invariant S-matrices at the four derivative order can be constructed from eq \eqref{tspinone4der}. This is a polynomial of $\xi$ with two possible choices for each particle. It is easy to see that two of the the possible $\Z_2 \times \Z_2$ combinations are  the structures having all $\alpha_i$s and all $\beta_i$s.  

\begin{equation}\label{04and40}
    \begin{split}
        & M^{\mathfrak{v}, (4,0)}_{\partial^4,1}=\alpha_1\alpha_2\alpha_3\alpha_4,\qquad  M^{\mathfrak{v}, (0,4)}_{\partial^4,1}=\beta_1\beta_2\beta_3\beta_4,
    \end{split}
\end{equation}
where the superscript $(a, b)$ denotes the number of $\alpha$ and $\beta$ s respectively. For structures having two $
\alpha_i$s and two $\beta_i$s the possibilities are three in number 

\begin{equation}\label{22and22}
    \begin{split}
        M^{\mathfrak{v}, (2,2)}_{\partial^4,1}=\alpha_1 \alpha_2 \beta_3 \beta_4|_{\Z_2\times \Z_2},\qquad M^{\mathfrak{v}, (2,2)}_{\partial^4,2}=\alpha_1 \beta_2 \alpha_3 \beta_4|_{\Z_2\times \Z_2},\qquad M^{\mathfrak{v}, (2,2)}_{\partial^4,3}=\alpha_1 \beta_2 \beta_3\alpha_4|_{\Z_2\times \Z_2}.
    \end{split}
\end{equation}
Similarly for structures having one $
\alpha_i$ and three $\beta_i$s and vice versa, 
\begin{equation}\label{13and31}
    \begin{split}
        M^{\mathfrak{v}, (1,3)}_{\partial^4,1}=\alpha_1 \beta_2 \beta_3 \beta_4|_{\Z_2\times \Z_2},\qquad M^{\mathfrak{v}, (3,1)}_{\partial^4,1}=\alpha_1 \alpha_2 \alpha_3 \beta_4|_{\Z_2\times \Z_2},
    \end{split}
\end{equation}
where these structures are to be considered after imposing $\Z_2\times \Z_2$. In total we have seven structures. Proceeding as in the two derivative case, we get the following $S_3$ decomposition of S-matrices,

\begin{equation}\label{fourders3spinone}
7 =  {\bf 1_S} + 3~ {\bf 2_M}.    
\end{equation}
In total therefore, we find nineteen quasi-invariant S-matrices for spin one scattering.

\section{Explicit construction of the S-matrices: spin two}\label{ecosms2}

We now evaluate independent scattering amplitudes relevant for massive spin two scattering. The massive spin two field admits plane wave solutions parametrized by a symmetric traceless tensor $h_{\mu \nu}(p_i)$. As explained previously, it is often convenient to specialise to a special choice of polarisation without any loss of generality, 

 \begin{eqnarray}
     h_{\mu \nu} (p_i)= \epsilon_\mu (p_i) \epsilon_\nu (p_i),\qquad \epsilon (p_i). \epsilon (p_i) =0,~~~ \epsilon (p_i). \epsilon^{*} (p_i) =1,
 \end{eqnarray}
where the second constraint implements the tracelessness condition and third condition is a normalisation condition. This enables us to classify the structures relevant to gravitational scattering in terms of building blocks eq \eqref{VarDef}.
The exact expressions for all the structures are recorded in the ancillary \texttt{Mathematica} file \textit{spintwo\_regge.nb}. Here we present the counting of the number of the structures, in terms of the scalar invariants without presenting the explicit expressions. We grade the structures by their derivatives. 

\subsection{Non-identical particles}

          We first enumerate the non-identical amplitudes. Similar method is followed, we list by counting structures which saturate homogeneity in polarisations as polynomials of $\xi_i$ and $b_{ij}$. Then estimate the possible degeneracies of each such structure because of two possible values of $\xi_i$.

\subsubsection*{Zero derivative S-matrices}

Structures at Zero derivatives are just polynomials of $b_{ij}$. The number of independent $b_{ij}$ structures are 6. We can think of it as the number of independent structures of $b_{1 \sigma(1)}b_{2\sigma(2)}b_{3 \sigma(3)}b_{4\sigma(4)}$ with the constraint, $b_{ii}=0$ because of the tracelessness of $h_{\mu\nu}$ and $b_{ij}= b_{ji}$ because of symmetricity of $h_{\mu\nu}$. It is helpful to think in terms of the conjugacy classes of the $S_4$ permutation group. Due to the condition $b_{ii}=0$, the two conjugacy classes which survive are the double transposition (or $\mathbb{Z}_2 \times \mathbb{Z}_2$) and the four cycles generated by, 
\begin{eqnarray}
(12)(34),~(13)(24),~(14)(23),\nonumber\\
(1234),~(1243),~(1324),~(1342),~(1423),~(1432).
\end{eqnarray}
The $\mathbb{Z}_2 \times \mathbb{Z}_2$ action generates terms which are quadratic in $b_{ij}$, these are three in number. We further note that due to the symmetricity of the scalar $b_{ij}$, three of the amplitudes generated by six four cycles are independent. Thus the number of linearly independent zero derivative structures are 6. In equations, we have two classes of zero derivative structures with three linearly independent structures for both of them,

\begin{equation} \label{tspintwozeroder}
    M^{\mathfrak{h}}_{\partial^0, 1}= b_{ij}^2 b_{kl}^2,\qquad M^{\mathfrak{h}}_{\partial^0, 2}=b_{ij} b_{jk} b_{lk}b_{li}, \qquad \forall~~ i,j,k,l= 1,\cdots 4 ,
\end{equation}
where the superscript denotes this is a structure corresponding to spin two scattering and the subscript denotes the derivative order of the structure and labels the degeneracy. It is also implied that $i\neq j$ in $b_{ij}$.  

\subsubsection*{Two derivative S-matrices}

We now systematically classify the higher derivative structures. At two derivatives we expect the following three classes of structures
\begin{equation}\label{tspintwotwoder}
    M^{\mathfrak{h}}_{\partial^2, 1}=\xi_i \xi_j b_{ij} b_{kl}^2,\qquad, M^{\mathfrak{h}}_{\partial^2, 2}=\xi_i \xi_j b_{il} b_{lk}b_{kj},\qquad M^{\mathfrak{h}}_{\partial^2, 3}=\xi_i^2  b_{jl} b_{lk}b_{kj}, 
\end{equation}
where it is implied that the indices $i,j,k$ and $l$ take values from 1 to 4. We now count each of such possible structures. For the structure $M^{\mathfrak{h}}_{\partial^2, 1}$, we can choose $\xi_i,\xi_j$ in ${4 \choose 2}$ ways. Note that the structure $b_{ij} b_{kl}^2$ is automatically fixed by this choice. Hence the total counting for $M^{\mathfrak{h}}_{\partial^2, 1}$ becomes ${4 \choose 2}\times 2^2$ (Recall that we had two choices for each $\xi_i$).  For the structure $M^{\mathfrak{h}}_{\partial^2, 2}$, we can similarly choose $\xi_i,\xi_j$ in ${4 \choose 2}$ ways. However for the tensor structure $b_{il} b_{lk}b_{kj}$, we have two possible choices due to the fact that the indices $l$ and $k$ take two possible values. In other words, once $i$ and $j$ are fixed, there are two possible chains of $b_{il} b_{lk}b_{kj}$ with two fixed end points- $b_{il} b_{lk}b_{kj}$ and $b_{ik} b_{lk}b_{lj}$. Hence the total counting for $M^{\mathfrak{h}}_{\partial^2, 2}$ becomes ${4 \choose 2}\times 2\times 2^2$. For the tensor structure $M^{\mathfrak{h}}_{\partial^2, 3}$, we can choose $\xi_i^2$ in $4$ ways but the possible choices for $\xi_i^2$ is now three (corresponding to $\alpha_i^2, \beta_i^2$ and $\alpha_i\beta_i$) and moreover, the tensor structure $b_{jl} b_{lk}b_{kj}$ is fixed. The counting becomes  ${4 \choose 1}\times 3$. In total, we get 84 structures. 

Alternatively, we can also construct $M^{\mathfrak{h}}_{\partial^2, 1},~~ M^{\mathfrak{h}}_{\partial^2, 2}$ from the zero derivative structures by replacing $b_{ij}$s by $\xi_i\xi_j$. For $M^{\mathfrak{h}}_{\partial^2, 2}$, consider the zero derivative non-quadratic structures, there are three structures and from each structure, we can choose 1 $b_{ij}$ in ${4 \choose 1}$ way and make it a scalar of type $\xi_i \xi_j$. So the total number of ways is ${4 \choose 1} \times 3 \times 2 \times 2 $. For $M^{\mathfrak{h}}_{\partial^2, 1}$, we get 2 ways to replace the  $b_{ij}$ by $\xi_i \xi_j$ and there are three such structures so we get $2\times 3 \times 2 \times 2$.

\subsubsection*{Four derivative S-matrices}
For four derivative structures, we have the following possible tensors, 
\begin{equation}\label{tspintwofourder}
     M^{\mathfrak{h}}_{\partial^4, 1}=\xi^2_i \xi^2_j b_{kl}^2,\qquad, M^{\mathfrak{h}}_{\partial^4, 2}=\xi_i^2 \xi_j\xi_l b_{jk} b_{kl},\qquad M^{\mathfrak{h}}_{\partial^4, 3}=\xi_i\xi_j\xi_k\xi_l  b_{ij} b_{kl}. 
\end{equation}
In a similar manner, we enumerate the degeneracy of each class. For $M^{\mathfrak{h}}_{\partial^4, 1}$, we can choose  $i$ and $j$ indices in 6 different ways and each $\xi_i^2$ can have three possibilities which brings our counting to $6\times 3^2$. For the second structure $M^{\mathfrak{h}}_{\partial^4, 2}$, we can choose $\xi_i^2$ in 4 possible ways, which leaves us with three possibilities to choose $\xi_j \xi_l$, which automatically fixes the rest of the tensor structure. Our counting becomes $(4 \times 3)\times (3 \times 2^2)$, where we have taken into account that $\xi_i^2$ can take three values while $\xi_i$ can take two values. For the second structure $M^{\mathfrak{h}}_{\partial^4, 3}$, we note that we can construct $b_{ij}b_{kl}$ in three different ways (This is essentially the zero derivative photon structure), which also automatically fixes the $\xi_i$s. Hence the counting becomes, $3\times 2^4$.  In total we have 246 structures at four derivatives. 

\subsubsection*{Six derivative S-matrices}
For six derivative structures, we have the following possible tensor, 
\begin{equation}\label{tspintwosixder}
     M^{\mathfrak{h}}_{\partial^6, 1}=\xi^2_i \xi^2_j \xi_k \xi_l b_{kl}. 
\end{equation}

Recall that we had 6 independent $b_{ij}$s. Homogeneity in the polarisations ensures that the rest of the structure is fixed once we make a choice for $b_{ij}$. Hence the counting becomes, $6\times 2^2 \times3^2$ giving us 216 structures at six derivatives.

\subsubsection*{Eight derivative S-matrices}

At 8 derivative, we can only have one possible tensor structure,

\begin{equation}\label{tspintwo8der}
    \begin{split}
        & M^{\mathfrak{h}}_{\partial^8, 1}= \xi_i^2\xi_j^2\xi_k^2\xi_l^2. \\
       \end{split}
\end{equation}
The counting becomes $ 3^4= 81$.

Hence, the total number of spin two structures is 633.

\subsection{Identical particles}\label{irrepsspin2}
In this section we enumerate the reduction in the number of four point scattering amplitudes for spin two scattering if the particles were identical. Analogous to the case of identical scattering of spin one particles, we impose $S_4$ invariance in two steps. We will enumerate the quasi-invariant structures and their $S_3$ irreps, relegating the explicit formulae for the amplitudes to our \texttt{Mathematica} file \textit{spintwo\_regge.nb}. 

\subsection*{Zero derivative quasi-invariant S-matrices}

There are six structures at zero derivatives given by eq \eqref{tspintwozeroder}. We immediately see that these two classes of structures are $\Z_2 \times \Z_2$ invariant by themselves. The generators of the local module $M^{\mathfrak{h}}_{\partial^0,1}$ and $M^{\mathfrak{h}}_{\partial^0,2}$ transform in ${\bf 3}= {\bf 1_S}+{\bf 2_M}$ which accounts for six quasi invariant structures. 
\begin{equation}
    6= 2~{\bf 1_S}+2~{\bf 2_M}.
\end{equation}

\subsection*{Two derivative quasi-invariant S-matrices}

There are three classes of two derivative structures contributing to massive spin two scattering given by eq  \eqref{tspintwotwoder}. We think of a subset of the graviton structures as the product of the two spin one modules at zero derivative and two derivative order, 
\begin{equation}
    \begin{split}
        M^{\mathfrak{v}}_{\partial^2, 1}|_{\Z_2 \times \Z_2}= \left(\xi_i \xi_j b_{kl} + \xi_k \xi_l b_{ij}\right), \qquad M^{\mathfrak{v}}_{\partial^0, 1}|_{\Z_2 \times \Z_2}= b_{ij} b_{kl} 
    \end{split}
\end{equation}The quasi-invariant structures that can be constructed from them are as follows. 

\begin{equation}
    \begin{split}
        &M^{\mathfrak{h}}_{\partial^2,1}|_{\Z_2 \times Z_2} = \left( \xi_i \xi_j b_{ij} b_{kl}^2+ \xi_k \xi_l b_{kl} b_{ij}^2\right),\qquad M^{\mathfrak{h}}_{\partial^2,2}|_{\Z_2 \times Z_2} = \left( \xi_i \xi_j b_{il} b_{lk}b_{kj}+ \xi_k \xi_l b_{il} b_{ij}b_{kj}\right),\\ 
    \end{split}
\end{equation}

For the structure in  $M^{\mathfrak{h}}_{\partial^2,1}|_{\Z_2 \times Z_2}$ we note that it is generated by the product of the spin one structures which remain invariant under the same $\Z_2 \times \Z_2$ cycle, while for the second structure, it is the product of $M^{\mathfrak{v}}_{\partial^2, 1}|_{\Z_2 \times \Z_2}$ with the other $\Z_2 \times \Z_2$ invariant element in the orbit of $M^{\mathfrak{v}}_{\partial^0, 1}|_{\Z_2 \times \Z_2}$. For the class of structures, $M^{\mathfrak{h}}_{\partial^2,1}|_{\Z_2 \times Z_2}$, we can choose the indices $i, j$ in $3$ ways because of $\Z_2 \times \Z_2$ symmetry which fixes rest of the structure and there are three $\Z_2$ symmetric (of $\Z_2\times \Z_2$ ) combinations  possible for $\xi_i \xi_j$. We therefore obtain $3 \times 3 =9$ for $M^{\mathfrak{h}}_{\partial^0, 1}|_{\Z_2 \times \Z_2}$. While for $M^{\mathfrak{h}}_{\partial^2,2}|_{\Z_2 \times Z_2}$, we can choose the indices $i, j$ in $3$ ways and there is a further choice of ${2 \choose 1}$ to be made for the tensor structure $b_{il} b_{lk}b_{kj}$ (because of the chain like structure). Therefore, we obtain $3\times3\times 2=18$, giving us a total of 27 quasi-invariant S-matrices for spin two. The third structure which cannot be expressed as product of two spin one structures is given by,
\begin{equation}
    \begin{split}
        M^{\mathfrak{h}}_{\partial^2,3}|_{\Z_2 \times Z_2} =\left(\xi_1^2 b_{23}b_{34}b_{24}+ \xi_2^2 b_{13}b_{34}b_{14}+\xi_3^2 b_{12}b_{24}b_{14}+\xi_4^2 b_{12}b_{23}b_{13}\right).
    \end{split}
\end{equation}
It is easy to see that we have three such structures corresponding to choices of $\xi_i^2$ (see eq \eqref{xisqposs}). So the total number of two derivative quasi invariant structures becomes 30 and their $S_3$ representation is given by, 
\begin{eqnarray}
 30=6~{\bf 1_S} + 4~{\bf 1_A} + 10~{\bf 2_M}. 
\end{eqnarray}
We derive this explicitly by using the $S_3$ projectors (see eq \eqref{s3proj}). We provide the explicit construction of the irreps in the \texttt{Mathematica} files.

\subsubsection*{ Four derivative quasi-invariant S-matrices }
The four derivative quasi-invariant S-matrices fall into the following three classes from eq \eqref{tspintwofourder}
\begin{equation}
\begin{split}
     &M^{\mathfrak{h}}_{\partial^4, 1}|_{\Z_2 \times \Z_2}=(\xi^2_i \xi^2_j b_{kl}^2+\xi^2_k \xi^2_l b_{ij}^2),\\
     & M^{\mathfrak{h}}_{\partial^4, 2}|_{\Z_2 \times \Z_2}=(\xi_i^2 \xi_j\xi_l b_{jk} b_{kl}+ \xi_j^2 \xi_i\xi_k b_{il} b_{kl}+\xi_k^2 \xi_l\xi_j b_{li} b_{ij}+\xi_l^2 \xi_i\xi_k b_{il} b_{lk}),\\
     &M^{\mathfrak{h}}_{\partial^4, 3}|_{\Z_2 \times \Z_2}=\xi_i\xi_j\xi_k\xi_l  b_{ij} b_{kl}.  
\end{split}
    \end{equation}
For the first class of quasi-invariant amplitudes, the number of independent amplitudes can be counting by noting that for all we have 3 possibilities for the $i, j$ indices due to $\Z_2 \times \Z_2$ symmetry. For each of them the product $\xi_i^2 \xi_j^2$ takes six possible combinations symmetric in $\Z_2$ of $\Z_2\times \Z_2$ in terms of $\alpha_i$ and $\beta_i$
\begin{equation}
\begin{split}
    &\alpha_i^2\alpha_j^2,~~\alpha_i^2\beta_j^2 +\alpha_j^2\beta_i^2,~~\alpha_i^2\alpha_j\beta_j +\alpha_j^2\alpha_i\beta_i,\\
    & \alpha_i \beta_i \alpha_j \beta_j,~~ \alpha_i \beta_i \beta_j^2+\alpha_j \beta_j \beta_i^2, ~~\beta_i^2 \beta_j^2.
\end{split}
    \end{equation}
Thus there are 18 structures in total for the first class. For the second class, we note that the primary structure from which this is generated, requires explicit $Z_2 \times \Z_2$ symmetrization. For a given index $i$, there are ${3 \choose 2}$ ways of choosing the indices $j$ and $k$ and this fixes the rest of the structure. Taking into the fact that $\xi_i^2$ and $\xi_i$ can take three and two possible values respectively, the number of structures is $3\times 3\times 2^2= 36$. For the third structure, the number of linearly independent $\Z_2 \times \Z_2$ invariant combinations possible is three (basically set by $b_{ij}b_{kl}$). The possible choices for the combination  $\xi_i\xi_j\xi_k\xi_l$, was evaluated in eq \eqref{fourders3spinone} giving a total of 21 structures of this class. The total number of four derivative quasi-invariant structures therefore becomes 75.

Through explicit use of the projectors outlined in the eq \eqref{s3proj}, we explicitly construct the $S_3$ representations of our quasi invariant S-matrices,

\begin{equation}
    75= 15~{\bf 1_S} + 25~{\bf 2_M} +10~{\bf 1_A}.
\end{equation}

\subsection*{Six derivative quasi-invariant S-matrices}
The possible class of quasi-invariant structures at six derivatives is given by,
\begin{equation}
     M^{\mathfrak{h}}_{\partial^6, 1}|_{\Z_2 \times \Z_2}=(\xi^2_i \xi^2_j \xi_k \xi_l b_{kl}+\xi^2_l \xi^2_k \xi_j \xi_i b_{ji}).
\end{equation}
The counting is facilitated by rewriting this as product of the four derivative and two derivative spin one module as follows, 

\begin{equation}
     M^{\mathfrak{h}}_{\partial^6, 1}|_{\Z_2 \times \Z_2}=(\xi_i\xi_j\xi_k\xi_l)(\xi_i \xi_j  b_{kl}+\xi_l \xi_k  b_{ji}).
\end{equation}
The term in the first bracket is $\Z_2 \times \Z_2$ invariant by itself with seven possibilities as explained before eq \eqref{fourders3spinone}. The term in the second bracket has nine possibilities as explained around eq \eqref{twoders3spinone}. In total there are 63 quasi-invariant S-matrices. Their $S_3$ properties can be worked out similarly, 

\begin{equation}
    63=11~{\bf 1_S} + 21~{\bf 2_M} + 10~{\bf 1_A}
\end{equation}

\subsubsection*{ Eight derivative quasi-invariant S-matrices }
For this case, it is convenient to write the tensor structure again in terms of product of two spin one four derivative structure, 
\begin{equation}
    \begin{split}
        & M^{\mathfrak{h}}_{\partial^8, 1}= (\xi_i\xi_j\xi_k\xi_l)(\xi_i\xi_j\xi_k\xi_l), \\
       \end{split}
\end{equation}
We can there fore construct these amplitudes explicitly using product of the spin one four derivative amplitudes eq \eqref{22and22}, eq \eqref{04and40} and eq \eqref{13and31}. Schematically the amplitudes are given by 

\begin{equation}
    M^{\mathfrak{v}, (\alpha,\beta)}_{\partial^4,i} M^{\mathfrak{v}, (\gamma,\delta)}_{\partial^4,j},\qquad \alpha+\beta=\gamma+\delta=4,\qquad i\leq j
\end{equation}
The number of such product amplitudes turn out to be 28 but with one relation amongst them, 

\begin{eqnarray}
\begin{split}
   M^{\mathfrak{v}, (0,4)}_{\partial^4,1}M^{\mathfrak{v}, (4,0)}_{\partial^4,1} &=-\frac{1}{4}\left(M^{\mathfrak{v}, (2,2)}_{\partial^4,1} M^{\mathfrak{v}, (2,2)}_{\partial^4,2}+M^{\mathfrak{v}, (2,2)}_{\partial^4,1} M^{\mathfrak{v}, (2,2)}_{\partial^4,3}+M^{\mathfrak{v}, (2,2)}_{\partial^4,2} M^{\mathfrak{v}, (2,2)}_{\partial^4,3}\right)\\
   &+\frac{1}{4}M^{\mathfrak{v}, (1,3)}_{\partial^4,1} M^{\mathfrak{v}, (3,1)}_{\partial^4,1}.  
\end{split}
   \end{eqnarray}
Thus, we find that there are 27 independent amplitudes. Their $S_3$ representations are worked out to be, 
\begin{equation}
    27=6~{\bf 1_S} +9~{\bf 2_M}+ 3~{\bf 1_A}.
\end{equation}

In total we find two hundred and one quasi-invariant S-matrices for spin two scattering.

\section{CRG allowed four point couplings}\label{CRGstruc}

 In this section we address the main question of our paper and study the Regge growth of our non-identical as well as identical amplitudes for spin one and spin two particles. Each of the amplitudes classified in \S\ref{ecosms1} is generated by unique four point coupling in the Lagrangian, which, in principle, can contribute to 23 different processes other than the one studied in \S\ref{ecosms1}, when the particle masses are different.  Classical Regge Growth states that four point tree-level scattering amplitudes should always grow slower that $s^2$ at large $s$ and fixed $t$ (where $s$ and $t$ are Mandelstam variables) for all values of the physical momenta and for all values of the polarisation tensors (if the external states have spin). The Lagrangians we classified in the previous sections occur at different orders in derivatives and the most general S-matrix can be schematically generated by the Lagrangian 
 \begin{equation}\label{lagbody}
     \sum_i \gamma_i L_i,
 \end{equation}
 where $\gamma_i$ are dimensionful couplings and $L_i$ are quartic Lagrangians which generate the primaries (or local module for identical scattering) and their descendants classified in the previous section. CRG states that for the tree-level scattering amplitude $T(s,t)$
generated by eq \eqref{lagbody},
\begin{eqnarray}
 \lim_{s \rightarrow \infty,~ \textrm{fixed} ~t} T(s,t) \leq s^2 ,\qquad   \frac{1}{\Lambda^2} \ll s\ll \frac{1}{l_p^2}, 
\end{eqnarray}
where $\Lambda$ is a low energy scale relevant to the theory we are studying. As elucidated in the introduction, $\Lambda$ can be the string length for tree-level string theory or the compactification length for the Kaluza Klein theory and $l_p$ is the Planck length.  We only consider the tree-level processes in such a theory, since loops in these examples would be suppressed by $l_p$. For non-identical particles, CRG implies that a linear combination of the classified interactions (as well as its descendants) has to obey CRG for each of the 24 processes. For identical particles, all the processes are the same and CRG then implies that linear combination of S-matrices obtained by $S_3$ projection of quasi-invariants must grow slower than $s^2$ for large $s$ at fixed $t$.

 \subsection{Massive spin one couplings}\label{CRGALLWDl1}
 We first demonstrate how to obtain the amplitude for different processes from a single process. Let us be precise and consider a processes generated due to the following spin one Lagrangian term, 

\begin{eqnarray}\label{testlag1}
    {A_1}_\mu \partial_\nu A_2^\mu {A_3}_\alpha \partial^\nu A_4^\alpha, 
\end{eqnarray}

where the spin one fields $A_1, A_2, A_3$ and $A_4$ denote spin one particles of  masses $m_1, m_2, m_3$ and $m_4$ respectively. The scattering amplitude generated by the Lagrangian eq \eqref{testlag1} in the process where particle of mass $m_1 ,m_2$ are incoming while $m_3, m_4$ are outgoing, is given by, 

\begin{equation}\label{tell1}
    \begin{split}
        T^{\mathfrak{v}}_{12 \rightarrow 34} \equiv T^{\mathfrak{v}}_{\epsilon_1(p_1)\epsilon_2(p_2) \rightarrow \epsilon_3(p_3)\epsilon_4(p_4)}=- \epsilon_1(p_1)\cdot \epsilon_2(p_2) \epsilon_3(p_3)\cdot \epsilon_4(p_4) (p_2\cdot p_4).
    \end{split}
\end{equation}

We also have to consider the process in which particles $m_1, m_3$ are incoming and $m_2, m_4$ are outgoing. This corresponds to the amplitude \begin{eqnarray}\label{13to24v2}
    T^{\mathfrak{v}}_{\epsilon_1(p_1)\epsilon_3(p_2) \rightarrow \epsilon_2(p_3)\epsilon_4(p_4)},
\end{eqnarray}
with the on-shell conditions 
\begin{eqnarray}
    &&\epsilon_i(p_i)\cdot p_i =0,\qquad p_i^2=-m_i^2, \qquad i=1,4,\nonumber\\
    && \epsilon_i(p_j)\cdot p_j =0,\qquad p_i^2=-m_j^2,\qquad i,j = 2,3,\qquad i\neq j.
\end{eqnarray}
The scattering amplitude generated by the Lagrangian eq \eqref{testlag1} for this process is given by, 
\begin{equation}\label{1324eg}
         T^{\mathfrak{v}}_{\epsilon_1(p_1)\epsilon_3(p_2) \rightarrow \epsilon_2(p_3)\epsilon_4(p_4)}=- \epsilon_1(p_1)\cdot \epsilon_2(p_3) \epsilon_3(p_2)\cdot \epsilon_4(p_4) (p_3\cdot p_4),
\end{equation}
which is distinct from eq \eqref{tell1}. We can get the amplitude in eq \eqref{1324eg} directly from eq \eqref{tell1},
\begin{eqnarray}
    T^{\mathfrak{v}}_{\epsilon_1(p_1)\epsilon_3(p_2) \rightarrow \epsilon_2(p_3)\epsilon_4(p_4)} = T^{\mathfrak{v}}_{12\rightarrow 34}\left(\epsilon_i(p_{\sigma(i)}),p_{\sigma(i)}\right)
    ,\qquad p_{\sigma(i)}^2=-m_i^2,
\end{eqnarray}
where, for the scattering process in eq \eqref{13to24v2}, $\sigma= P_{23}$. In total, there could be twenty four different processes from the Lagrangian eq \eqref{testlag1}, which can be related by elements of $S_4$ to a single amplitude. Hence, we generate amplitudes for all the twenty processes from the process in which $m_1$ and $m_2$ are incoming and $m_3$ and $m_4$ are outgoing,
\begin{eqnarray}\label{tell11}
    T^{\mathfrak{v}}_{\sigma(12 \rightarrow 34)} \equiv T^{\mathfrak{v}}_{\epsilon_{\sigma(1)}(p_1)\epsilon_{\sigma(2)}(p_2) \rightarrow \epsilon_{\sigma(3)}(p_3)\epsilon_{\sigma(4)}(p_4)}= T^{\mathfrak{v}}_{12 \rightarrow 34}\left(\epsilon_i(p_{\sigma(i)}),p_{\sigma(i)}\right),~~~ p_{\sigma(i)}^2=-m_i^2,\nonumber\\
\end{eqnarray}
where $\sigma$ on the LHS acts on the mass labels without changing the incoming or outgoing momenta labels. However, solving for $\epsilon_i(p_{\sigma(i)})$ is a tedious process and an equivalent way (and also computationally convenient way) of getting the amplitudes for different processes is the following,  
\begin{eqnarray}\label{crossingregge}
    T^{\mathfrak{v}}_{\sigma(12 \rightarrow 34)} = T^{\mathfrak{v}}_{12 \rightarrow 34}\left(\epsilon_{\sigma(i)}(p_{\sigma(i)}), p_{\sigma(i)}\right) \bigg|_{m_{\bar{\sigma}(i)},~\polo_{\bar{\sigma}(i)}}, 
\end{eqnarray}
which implies we get the amplitude in two steps. First we permute the $\epsilon_i (p_i)$s in the process eq \eqref{12to34scat1} according to the relevant element of $S_4$ and then in this result {\it undo} the effect of permutation $\sigma$ in the masses and $\polo_i$. This enables us to solve the $\polp_a$ only once and hence deriving the amplitude for different processes computationally less arduous.

 We have classified in  \S\ref{ecosms1} and \S\ref{ecosms2}, all the contact scattering amplitudes for the particular process in eq \eqref{12to34scat} corresponding to massive spin one and two external states. Each of the scattering amplitudes is equivalent to a local Lagrangian. Our analysis implies that each of these different processes can be obtained directly from the scattering process in eq \eqref{12to34scat}, without explicitly deriving the Lagrangian interaction term generating it,
\begin{equation}\label{telljsigma}
  T^{\mathfrak{v}/\mathfrak{h}}_{j,~ \sigma(12 \rightarrow 34)} = T^{\mathfrak{v}/\mathfrak{h}}_{j,~ 12 \rightarrow 34}\left(\epsilon_i(p_{\sigma(i)}),p_{\sigma(i)}\right), \qquad p_{\sigma(i)}^2=-m_i^2,
    \end{equation}
where $\sigma$ denotes one of the elements of $S_4$ and $j$ keeps track of the original scattering amplitude or the Lagrangian giving rise to that amplitude in the process eq \eqref{12to34scat}. More precisely, for spin one, $j$ takes values from 1 to 43  while for massive spin two $j$ takes values from  1 to 633. Demanding that the amplitudes obey CRG in each of the twenty four processes, we arrive at the bounds on the classified scattering amplitudes. Using our kinematics described in \S\ref{kinematics}, we present the results below.

\subsubsection{Non-identical scattering}\label{ReggeSpin1}

The space of scattering amplitudes for the process $T^{\mathfrak v}_{12\rightarrow 34}$ has been evaluated in \S\ref{ecosms1} and  \S\ref{ecosms2}.  Explicit expressions for the same have been given in the ancillary \texttt{Mathematica} file \textit{spinone\_regge.nb} and \textit{spintwo\_regge.nb}. For spin one we have forty three structures labelled as $T^{\mathfrak{v}}_{j,~12\rightarrow 34}(\epsilon_i(p_i))$, where $j=1\cdots 43$. In general, all of these amplitudes are generated by local Lagrangians of the form 
\begin{equation}\label{spinonetwolag}
    \mathcal{L}= \sum_i \gamma_i L_i,
\end{equation}
where $i=1,\cdots 43$ for spin one and $\gamma_i$ are dimensionful parameters.  There is a one to one map (upto equations of motion and total derivatives) between the local Lagrangians $L_i$ and the classified amplitudes generated by the process $T^{\mathfrak v}_{12\rightarrow 34}$. However we will not explicitly derive the map in this work and its enough for the purpose of this note to know that such a unique map exists. Thus a generic four point scattering amplitude for spin one or spin two external states takes the form,
\begin{equation}
   T^{\mathfrak{v}/\mathfrak{h}}_{12\rightarrow 34,~total}= \sum_j \gamma_j T^{\mathfrak{v}/\mathfrak{h}}_{j,~12\rightarrow 34},
\end{equation}
where, on the RHS, $j=1,\cdots 43$ for spin one (while for spin two, $j=1,\cdots 633$) and the subscript $12\rightarrow 34$ in the amplitude $T^{\ell}_{j,~12\rightarrow 34}$ denotes the corresponding process. In order to classify the Regge allowed S-matrices, we search for linear combinations which are CRG allowed i.e, grow less than $s^2$ in the Regge limit. In particular, this linear combination has to be CRG allowed for {\it all} of the twenty four possible processes that we can have for non-identical particles.\begin{equation}
    T^{\mathfrak{v}/\mathfrak{h}}_{\sigma(12\rightarrow 34),~total}= \sum_j \gamma_j T^{\mathfrak{v}/\mathfrak{h}}_{j,~\sigma(12\rightarrow 34)},
\end{equation}where $T^{\mathfrak{v}/\mathfrak{h}}_{j,~\sigma(12\rightarrow 34)}$ is defined in eqs \eqref{tell11} and \eqref{telljsigma}. Note in particular, the coefficients $\gamma_j$  do not change since they label the Lagrangians rather than the process.

We now outline the algorithm to find the CRG allowed combination of couplings for spin one. Algebraically the criteria for finding the linear combination of $\gamma_i$s such that these amplitudes obey CRG, translates to a set of twenty four simultaneous equations in forty three couplings $\gamma_i$,\begin{equation}
\lim_{s \rightarrow \infty,~\textrm{fixed}~ t} T^{\mathfrak{v}}_{\sigma(12\rightarrow 34),~ total} \leq s^2,\qquad \forall~~ \sigma.
\end{equation} We first use eq \eqref{telljsigma} to generate the amplitudes for rest of 23 different processes obtained from the Lagrangians dual to each of the forty three amplitudes for the process in eq \eqref{12to34scat1}. Secondly, using the explicit parametrization of $\polp_i$s in \S \ref{kinematics} (more precisely eqs  \eqref{spinonepol} and \eqref{polpparaspinone}), we evaluate the amplitudes and expand $T^{\mathfrak{v}}_{\sigma(12\rightarrow 34),~ total}$ at large $s$ and fixed $t$. This expansion is organised as a power series expansion in $s^n$. At each order in $s^n$, the expansion contains polynomials of  the independent scattering parameters $\alpha_i, \beta_i$, $\polo_i$, polynomials of t such that the homogeneity is satisfied and we solve for the constraints on $\gamma_i$ by setting the coefficient of each polynomial of $\alpha_i, \beta_j, \polo_i\cdot\polo_j$ and $t$ (since these are independent data)to zero for $n=3,4$ . In particular, we allow for the possibility of mass dependent constraints among  the dimensionful parameters $\gamma_i$.  We explicitly demonstrate these steps in our ancillary \texttt{Mathematica} notebook \textit{spinone\_regge.nb}.

We just provide the counting of the final results here since the number of structures are too large and explicit expressions are un illuminating.  We get three Regge allowed structures in the zero-derivative order, twenty one Regge allowed structures in the second derivative order, and  two Regge allowed structures in the four-derivative order from the list $T^{\mathfrak{v}}_{j,1}(\epsilon_i(p_i))$. These structures imply we have twenty six local Lagrangians which are CRG allowed for any of the twenty four different scattering processes we study. We also find that constraints do not depend on mass. We can also consider descendants of these structures. Considering descendants leads to a larger space of allowed structures at each derivative order because of cancellations between the primary structures at each derivative order and descendants of lower primaries. It is a tedious but a finite process and we find the end results not illuminating enough to record them.

\subsubsection{Identical Scattering}\label{idscatspin1}

In this section we study the scattering of four external spin one states of the same mass $m$. For identical particles, there is no distinction between different processes. The amplitudes enjoy additional $S_4$ symmetry. Naively one would have thought that given the classification of CRG allowed non-identical amplitudes that we have already done, the allowed identical amplitudes would be the singlet projection of them. A crucial difference in these two cases is the fact that for the non-identical case, we demanded that the Regge growth be less than $s^2$ for each individual process which translates to each individual orbit of $S_4$. However, for the identical scattering, this is too strong a criteria and would exclude potential cancellations in the leading Regge behaviour between different orbits of $S_4$. In other words the $S_4$ invariance and the Regge limit do not commute. The correct order of limits is to impose $S_4$ invariance and then take Regge limit. 

We recall the classification into the irreps of $S_4$ that we had done for the massive spin one scattering amplitudes and summarise the classification in table \ref{tableqinvspin1} and provide explicit expressions for the module generators in the file \textit{spinone\_regge.nb}. \begin{table}
\begin{center}
\begin{tabular}{ |c|c|c|c|}
\hline
 $\partial^n$ &${\bf 1_S}$ &${\bf 2_M}$ & ${\bf 1_A}$ \\
 \hline 
 $n=4$ &1 & 3 & 0 \\
 \hline
 $n=2$ &2 & 3 & 1 \\
 \hline
 $n=0$ &1 & 1 & 0 \\
 \hline    
\end{tabular}
\end{center}
\caption{$S_3$ classification of spin one quasi-invariant modules (primaries) for $D\geq 8$ organised by derivatives, where $n$ denotes the derivative order of the structures.}\label{tableqinvspin1}
\end{table}We scan over the space of S-matrices order by order in derivatives. At each derivative order, we can construct the most general S-matrix (including descendants) using eq \eqref{mostgendescl1} and table \ref{tableqinvspin1}. Let us illustrate this construction in detail for spin one with the understanding that that the spin two construction, later on, will be done in the same manner.
The most general S-matrix at zero order in derivatives is generated by $e_{\bf 1_S}^{\partial^0}$. From table \ref{tableqinvspin1}, we see that at this derivative order we have only one such structure. At second order in derivatives, the basis at second order in derivatives is spanned by 
\begin{equation}
\begin{split}
\mathcal{B}^{\partial^2}_{\ell=1}\equiv\left(e_{\bf 1_S}^{\partial^2} \oplus \left((2s-t-u)e^{\partial^0,(1)}_{{\bf 2_M}}+\text{permutations}\right) \oplus m^2e_{\bf 1_S}^{\partial^0} \right),
\end{split}
\end{equation}
where we have considered the ``m"-descendant of the lower order guys. In total, at second order in derivatives, we find that there are four linearly independent S-matrices out of which two are primaries and two are descendants. We can similarly list the basis at fourth and sixth order in derivatives, 
\begin{equation}
\begin{split}
\mathcal{B}^{\partial^4}_{\ell=1}\equiv&\left(e_{\bf 1_S}^{\partial^4}\oplus (s^2+t^2+u^2)e^{\partial^0}_{{\bf 1_S}} \oplus \left((2s-t-u)e^{\partial^2,(1)}_{{\bf 2_M}}+\text{permutations}\right)\right.\\
&\left. \oplus\left((2s^2-t^2-u^2)e^{\partial^0,(1)}_{{\bf 2_M}}+\text{permutations}\right)   \oplus m^2 \mathcal{B}^{\partial^2}_{\ell=1} \right),\\
\mathcal{B}^{\partial^6}_{\ell=1}\equiv&\left( (s^2+t^2+u^2)e^{\partial^2}_{{\bf 1_S}} \oplus (stu)e^{\partial^0}_{{\bf 1_S}}\oplus \left((2s-t-u)e^{\partial^4,(1)}_{{\bf 2_M}}+\text{permutations}\right)\right.\\
&\left. \oplus\left((2s^2-t^2-u^2)e^{\partial^2,(1)}_{{\bf 2_M}}+\text{permutations}\right) \oplus \left(s^2+t^2+u^2\right)\left((2s-t-u)e^{\partial^0,(1)}_{{\bf 2_M}}+\text{permutations}\right) \right.\\
&\left.\oplus m^2 \mathcal{B}^{\partial^4}_{\ell=1} \right).
\end{split}
\end{equation}
From table \ref{tableqinvspin1}, we get that there are one primary and nine descendant structures at four derivatives and twenty descendant structures at six derivatives. Higher order descendants can similarly be constructed. The counting of S-matrices at each derivative order can be recast into a partition function which encodes the number of linearly dependent S-matrices at each derivative order and also their $S_3$ transformation properties \cite{Henning:2017fpj, Sundborg:1999ue, Aharony:2003sx, Chowdhury:2019kaq}. However, we will not explore that avenue in this present work for massive spinning particles.     

Having classified the basis at each derivative order, we can now summarise the Regge growth at each order in table \ref{CRGalwdspin1}. \begin{table}
	\begin{center}
		\begin{tabular}{ |c|c|c|}
			\hline
			$\partial^n$ & $\ell=1$ &$\ell=2$\\
			\hline 
			$n=6$ & 2  &1\\
			\hline
			$n=4$ & 4  &1\\
			\hline
			$n=2$ &3 &1\\
			\hline 
			$n=0$ &1 & 1\\
			\hline    
		\end{tabular}
	\end{center}
	\caption{Number of CRG allowed four point amplitudes for massive spin one and massive spin two identical scattering where $n$ denotes the derivative order}\label{CRGalwdspin1}
\end{table}
For this exercise, we use the explicit form of the polarisation vectors $\polp_i$s, obtained from eq \eqref{polpparaspinone} in the limit of equal external masses,

\begin{equation}\label{idpolparamet}
\begin{split}
\tilde{\epsilon}^L_{1} &= \tilde{\mathcal{N}}_1\left(\frac{k_1}{m} +\tilde{\mathcal{C}}_1 \frac{k_2}{m}+\tilde{\mathcal{C}}_2\frac{k_4}{m} \right),\qquad 
\tilde{\epsilon}^T_{1} = \tilde{\mathcal{N}}_2 \left(\tilde{\mathcal{T}}_1\frac{k_2}{m}-\tilde{\mathcal{T}}_2\frac{k_4}{m}\right),\\
\end{split}
\end{equation}
where, 
\begin{equation}
\begin{split}
&\tilde{\mathcal{N}}_1 =\sqrt{\frac{\tilde{s}^2-2 \tilde{s} \tilde{t}\tilde{u}+\tilde{u}^2}{\tilde{s}^2-2\tilde{s}\tilde{t}\tilde{u}+\tilde{t}^2+\tilde{u}^2-1}}, \quad \tilde{\mathcal{N}}_2= \sqrt{-\frac{\tilde{s}^2\tilde{u}^2}{\tilde{s}^2-2\tilde{s}\tilde{t}\tilde{u}+\tilde{u}^2}},\\ 
&\tilde{\mathcal{C}}_1 = \frac{(\tilde{u}\tilde{t}-\tilde{s})}{ \left(\tilde{s}^2-2\tilde{s}\tilde{t}\tilde{u}+\tilde{u}^2\right)}, \quad \tilde{\mathcal{C}}_2 = \frac{(\tilde{s} \tilde{t}-\tilde{u})}{\left(\tilde{s}^2-2\tilde{s}\tilde{t}\tilde{u}+\tilde{u}^2\right)}, \quad \tilde{\mathcal{T}}_1 = \frac{1}{\tilde{s}}, \quad \tilde{\mathcal{T}}_2 = \frac{1}{\tilde{u}},\\
&\tilde{s} = \frac{s}{2 m^2} - 1, \tilde{t}=\frac{t}{2 m^2} - 1 ~~\mathrm{ and}~~ \tilde{u} = \frac{u}{2 m^2} - 1,
\end{split}
\end{equation}
where $s$, $t$ and $u$ are the Mandelstam variables defined in eq \eqref{stu12to34} for equal masses. The rest of the $\polp_i$s are obtained as usual using $\Z_2 \times \Z_2$ projections of $\polp_1$. We have checked that there are no CRG allowed structures at eight derivatives which are not trivial $m$ descendants of the lower order structures. The explicit structures are cumbersome and we list them in the ancillary \texttt{Mathematica} file \textit{spinone\_regge.nb}. In the main text we provide a more concise form for the amplitude in terms of local Lagrangians. We find that the Regge allowed structures are generated by the following set of local Lagrangians,   
   
\begin{equation}\label{spin1crgalwdlag}
\begin{split}
\tilde{\mathcal{L}}_{\partial^0, 1}^{\mathfrak{v}} &= (A_\mu A^\mu)^2\\
\tilde{\mathcal{L}}_{\partial^2, 1}^{\mathfrak{v}} &=\partial_\sigma A_\mu A^\mu \partial^\sigma A_\nu A^\nu, \qquad\tilde{\mathcal{L}}_{\partial^2, 2}^{\mathfrak{v}} = \tilde{F}_{\alpha\beta}\tilde{F}^{\alpha\beta} A_\nu A^\nu,\qquad \tilde{\mathcal{L}}_{\partial^2, 3}^{\mathfrak{v}}= A^\mu F_{\mu \nu} \tilde{F}^{\nu \alpha} A_\alpha\\
\tilde{\mathcal{L}}_{\partial^4, 1}^{\mathfrak{v}} &= F_{\alpha \beta}F^{\beta \gamma}F_{\gamma \delta }F^{\delta \alpha},\qquad \tilde{\mathcal{L}}_{\partial^4, 2}^{\mathfrak{v}} = (F_{\alpha\beta}F^{\alpha\beta})^2, \qquad \tilde{\mathcal{L}}_{\partial^4, 3}^{\mathfrak{v}}= F_{ab}\tilde{F}^{bc}F_{cd}\tilde{F}^{da},\\
\tilde{\mathcal{L}}_{\partial^4, 4}^{\mathfrak{v}} &=A_{\mu} \partial_\sigma A_\nu \partial^\mu \partial^\sigma A_\alpha \partial^\nu A^\alpha,\\
\tilde{\mathcal{L}}_{\partial^6, 1}^{\mathfrak{v}} &= \partial_\sigma F_{\alpha\beta}F^{\beta \gamma }\partial^\sigma F_{\gamma \delta }F^{\delta \alpha}, \qquad \tilde{\mathcal{L}}_{\partial^6, 2}^{\mathfrak{v}} =  F_{\alpha\beta} \partial^\alpha F^{\gamma \delta}\partial^\beta F_{\delta \eta}F^{\eta \gamma},
\end{split}
\end{equation}

where we have defined $F_{ab}= \partial_a A_b-\partial_b A_a$ and $\tilde{F}_{ab}= \partial_a A_b+\partial_b A_a$. In the labelling of the structures, the superscript denotes it generates a massive spin one amplitude while the superscripts keep track of derivative order and the number of structures at each derivative order. In particular we note that the structures $\tilde{\mathcal{L}}_{\partial^4, 1}^{\mathfrak{v}},  \tilde{\mathcal{L}}_{\partial^4, 2}^{\mathfrak{v}}, \tilde{\mathcal{L}}_{\partial^6, 1}^{\mathfrak{v}}$ and $\tilde{\mathcal{L}}_{\partial^6, 2}^{\mathfrak{v}}$ were noted in \cite{Chowdhury:2019kaq} as Regge allowed structures for massless spin one scattering as well. The appearance of these polynomials for the massive case as well can be justified as follows. For massless spin one particles, since we have gauge invariance, the local Lagrangians are functions of the field strength. For massive spin one, as we have seen that we do not have this constraint and this leads to a broader class of Lagrangians which are not strictly polynomials in field strength. However the gauge invariant combination of field strength ensures that the longitudinal component (i.e the fastest growing component for the massive polarisation) of the spin one polarisation cancel and the Regge limit of polynomials of field strength evaluated over solutions of massive polarisations grow similarly as massless ones. Hence the space of Regge allowed structures for the identical massive spin one scattering should atleast include the Regge allowed structures corresponding to the massless spin one case and possibly some more. Happily we find this indeed is the case and this serves as a sanity check for our bootstrap method and the Regge analysis.

\subsection{Massive spin two couplings}\label{CRGALLOWED}

In this section we perform the same exercise for massive spin two scattering amplitudes. We analyse both non-identical and identical scattering in these subsequent subsections.

\subsubsection{Non-identical scattering}

As for the case of massive spin one particles, we use eq \eqref{telljsigma} to generate the amplitudes for rest of 23 different processes obtained from the Lagrangians dual to each of the six hundred thirty three amplitudes (evaluated in \S \ref{ecosms2}) for the process in eq \eqref{12to34scat1}. Secondly, using the explicit parametrization of $\polp_i$s in \S \ref{kinematics} (more precisely eqs  \eqref{spinonepol} and \eqref{polpparaspintwo2}), we do the Regge expansion of the amplitude in $s^n$. We solve for the constraints on $\gamma_i$ by setting the coefficient of each polynomial of $\alpha_i, \beta_j, \polo_i\cdot\polo_j$ and $t$ (since these are independent data) in the expansion to zero for $n$ which now runs from 3 to 8. We find five Regge allowed S-matrices for the process $T^{\mathfrak{h}}_{12 \rightarrow 34}$ ( and also for 23 different processes,  $T^{\mathfrak{h}}_{\sigma(12 \rightarrow 34)}$) which are listed below.

 \begin{equation}\label{crgallowedspin2nonid}
    \begin{split}
        \mathcal{A}_1 =& b_{14}^2 b_{23}^2-2 b_{14} \left(b_{13} b_{24}+b_{12} b_{34}\right) b_{23}+\left(b_{13} b_{24}-b_{12} b_{34}\right){}^2\\
        \mathcal{A}_2 =& A_{43}^2 b_{12} b_{13} b_{23}+A_{43} \left[\left(b_{13} b_{24}-b_{14} b_{23}\right) \left(A_{23} b_{13}-A_{13} b_{23}\right)-b_{12} b_{34} \left(A_{13} b_{23}+A_{23} b_{13}\right)\right]\\
        &+b_{34} \left(\left(A_{23} b_{13}-A_{13} b_{23}\right) \left(A_{23} b_{14}-A_{13} b_{24}\right)+A_{13} A_{23} b_{12} b_{34}\right),\\
        \mathcal{A}_3 =& A_{32}^2 b_{12} b_{14} b_{24}- A_{21} A_{32} b_{14}^2 b_{23}+ 2 A_{34} A_{32} b_{12} b_{14} b_{24}+ A_{21} A_{32} b_{13} b_{14} b_{24}- A_{14} A_{34} b_{13} b_{24}^2\\
        &- A_{14} A_{24} b_{12} b_{34}^2+A_{23} A_{34} b_{14}^2 b_{23}+A_{41}^2 b_{12} b_{13} b_{23}-A_{23} A_{34} b_{13} b_{14} b_{24}+A_{14} A_{34} b_{14} b_{23} b_{24}\\
        &+b_{34} \left[b_{14} \left(A_{21} \left(A_{32} b_{12}-2 A_{23} b_{13}-A_{14} b_{23}\right)-A_{23} \left(A_{34} b_{12}+A_{23} b_{13}+A_{14} b_{23}\right)\right)\right.\\
        &\left.-A_{14} b_{24} \left(-A_{34} b_{12}-A_{24} b_{13}+A_{14} b_{23}\right)\right]+A_{41} \left(A_{21} b_{13} \left(-b_{14} b_{23}+b_{13} b_{24}-b_{12} b_{34}\right)\right.\\
        &\left.+A_{32} b_{12} \left(b_{14} b_{23}+b_{13} b_{24}-b_{12} b_{34}\right)+A_{34} b_{12} \left(b_{14} b_{23}+b_{13} b_{24}-b_{12} b_{34}\right)\right),\\
        \mathcal{A}_4 =& A_{21}^2 b_{13} b_{14} b_{34}+A_{21} \left[A_{34} b_{14} \left(-b_{14} b_{23}+b_{13} b_{24}+b_{12} b_{34}\right)+ 2 b_{34} A_{23} b_{13} b_{14}\right.\\
        &\left.+A_{14} b_{34} \left(b_{14} b_{23}+b_{13} b_{24}-b_{12} b_{34}\right)\right]-A_{14} A_{23} b_{12} b_{34}^2-A_{23} A_{34}  b_{23} b_{14}^2+A_{14} A_{34}  b_{13} b_{24}^2\\
        &+A_{34} b_{24} b_{14} \left(A_{34} b_{12}+A_{23} b_{13}-A_{14} b_{23}\right) +A_{23} b_{14} b_{34} \left(A_{34} b_{12}+A_{23} b_{13}+A_{14} b_{23}\right)\\
        &+A_{14} b_{24} b_{34} \left(-A_{34} b_{12}+A_{23} b_{13}+A_{14} b_{23}\right),\\
        \mathcal{A}_5 =& \left(A_{12} A_{23}-A_{14} A_{21}\right) b_{12} b_{34}^2+b_{34} \left(A_{14} \left(b_{23} \left(A_{21} b_{14}-A_{43} b_{12}\right)+b_{24} \left(-A_{34} b_{12}+A_{21} b_{13}-2 A_{12} b_{23}\right)\right)\right.\\
        &\left.+A_{23} \left(A_{43} b_{12} b_{13}-A_{12} b_{24} b_{13}+b_{14} \left(A_{34} b_{12}+2 A_{21} b_{13}-A_{12} b_{23}\right)\right)\right)\\
        &-\left(b_{13} b_{24}-b_{14} b_{23}\right) \left(A_{23} \left(A_{43} b_{13}-A_{34} b_{14}\right)+A_{14} \left(A_{43} b_{23}-A_{34} b_{24}\right)\right)\\
        &+A_{41} \left(2 A_{43} b_{12} b_{13} b_{23}-\left(b_{13} b_{24}-b_{14} b_{23}\right) \left(A_{21} b_{13}-A_{12} b_{23}\right)+b_{12} b_{34} \left(A_{21} b_{13}+A_{12} b_{23}\right)\right.\\
        &\left.+A_{34} b_{12} \left(-b_{14} b_{23}-b_{13} b_{24}+b_{12} b_{34}\right)\right)+ A_{32}A_{43} b_{12} \left(b_{14} b_{23}+b_{13} b_{24}-b_{12} b_{34}\right)\\
        &+A_{32}A_{21} b_{14} \left(b_{14} b_{23}-b_{13} b_{24}-b_{12} b_{34}\right)+A_{32}b_{24} \left(-2 A_{34} b_{12} b_{14}-A_{12} \left(b_{14} b_{23}-b_{13} b_{24}+b_{12} b_{34}\right)\right)
    \end{split}
\end{equation}

These amplitudes are generated by the local Lagrangians  in the process $T^{\mathfrak{v}}_{12 \rightarrow 34}$. 
The analysis in the section establishes that these Lagrangians also obey CRG also in {\it any} of the 23 other distinct processes that we can study for non-identical scattering.

\subsubsection{Identical scattering}\label{idscatspin2}

In this section we compute the Regge allowed  couplings contributing to polynomial S-matrices for massive spin two scattering. The procedure is same as outlined in \S \ref{idscatspin1}. We construct the space of most general S-matrices (i.e including primaries and descendants) using the data from table \ref{tableqinvspin2} and eq \eqref{mostgendescl1}. Following the steps exactly as in the case of spin one, we find that there are exactly one CRG allowed structure at zero, two, four and six derivatives and no higher derivative order structures. We present the explicit amplitudes for these Regge allowed structures in the \texttt{Mathematica} file \textit{spintwo\_regge.nb} and present here the local Lagrangians giving rise to them.

\begin{equation}\label{basisspin2id}
    \begin{split}
    \tilde{\mathcal{L}}_{\partial^0}^{\mathfrak{h}} &=  \,\, \delta_{[\alpha}^{\gamma}\delta_{\beta}^{\zeta}\delta_{\xi}^{\rho}\delta_{\delta]}^{\sigma}\,\,  h_\gamma^{\phantom{a}\alpha} \, h_\zeta^{\phantom{a}\beta}\,  h_\rho^{\phantom{a}\zeta}\,h_\sigma^{\phantom{a}\delta}, \\
     \tilde{\mathcal{L}}_{\partial^2}^{\mathfrak{h}} &=  \,\, \delta_{[\alpha}^{\gamma}\delta_{\beta}^{\zeta}\delta_{\xi}^{\rho}\delta_{\delta}^{\sigma}\delta_{\mu]}^{\nu}\,\, \partial_\nu \partial^\mu h_\gamma^{\phantom{a}\alpha} \, h_\zeta^{\phantom{a}\beta}\,  h_\rho^{\phantom{a}\zeta}\,h_\sigma^{\phantom{a}\delta}, \\
    \tilde{\mathcal{L}}_{\partial^4}^{\mathfrak{h}} &=  \,\, \delta_{[\alpha}^{\gamma}\delta_{\beta}^{\zeta}\delta_{\xi}^{\rho}\delta_{\delta}^{\sigma}\delta_{\mu}^{\nu}\delta_{\mu']}^{\nu'}\,\, \partial_\nu \partial^\mu h_\gamma^{\phantom{a}\alpha} \, \partial_{\nu'} \partial^{\mu'} h_\zeta^{\phantom{a}\beta}\,  h_\rho^{\phantom{a}\zeta}\,h_\sigma^{\phantom{a}\delta}, \\
    \tilde{\mathcal{L}}_{\partial^6}^{\mathfrak{h}} &=    \,\, \delta_{[\alpha}^{\gamma}\delta_{\beta}^{\zeta}\delta_{\xi}^{\rho}\delta_{\delta}^{\sigma}\delta_{\mu}^{\nu}\delta_{\mu'}^{\nu'}\delta_{\mu'']}^{\nu''}\,\, \partial_\nu \partial^\mu h_\gamma^{\phantom{a}\alpha} \, \partial_{\nu'} \partial^{\mu'} h_\zeta^{\phantom{a}\beta}\,  \partial_{\nu''} \partial^{\mu''}h_\rho^{\phantom{a}\zeta}\,h_\sigma^{\phantom{a}\delta}, \\
    \end{split}
\end{equation}

where $\delta_{[\alpha_1}^{\mu_1}\delta_{\alpha_2}^{\mu_2}\cdots\delta_{\alpha_n]}^{\mu_n}$ is concise expression for product of two Levi-Civita tensors $\varepsilon^{\mu_1\mu_2\cdots \mu_n}\varepsilon_{\alpha_1\alpha_2\cdots \alpha_n}$. Note that the Lagrangian which is of the order six derivatives reminds us of the second lovelock term that was classified as Regge allowed for massless graviton scattering in \cite{Chowdhury:2019kaq}. Indeed the S-matrix generated by this term involves a seven dimensional Levi-Civita tensor and is similar to the massless case,
\begin{equation} \label{Smatrix6derspin2} 
S_{\partial^6}=\left( \epsilon_1 \wedge \epsilon_2 \wedge \epsilon_3  \wedge 
\epsilon_4 \wedge k_1 \wedge k_2 \wedge k_3 \right)^2+ \text{$S_3$ permutations}.
\end{equation} 
Similar to the massive spin one case, we note that the second lovelock term was precisely the gauge invariant Lagrangian that was not ruled out by CRG in \cite{Chowdhury:2019kaq} for $D>6$. The four derivative lovelock like structure  gives a S-matrix involving six dimensional Levi-Civita tensor of the form,
\begin{equation} \label{Smatrix4derspin2} 
S_{\partial^4}=\left( \epsilon_1 \wedge \epsilon_2 \wedge \epsilon_3  \wedge 
\epsilon_4 \wedge k_1 \wedge k_2  \right)^2 + \left( \epsilon_1 \wedge \epsilon_2 \wedge \epsilon_3  \wedge 
\epsilon_4 \wedge k_3 \wedge k_4  \right)^2  + \text{$S_3$ permutations}.
\end{equation} 
Since gauge invariance is no longer a requirement for massive spinning particles, we expect such structures to appear. Similarly, the two and zero derivative structures arise from ``lovelock-like"  Lagrangians which generate S-matrices involving five dimensional and four dimensional Levi-Civita tensors of the form,
\begin{equation} \label{Smatrix2derspin2}
\begin{split}
S_{\partial^2}=&\left( \epsilon_1 \wedge \epsilon_2 \wedge \epsilon_3  \wedge 
\epsilon_4 \wedge k_1  \right)^2 + \left( \epsilon_1 \wedge \epsilon_2 \wedge \epsilon_3  \wedge 
\epsilon_4 \wedge k_2  \right)^2+ \left( \epsilon_1 \wedge \epsilon_2 \wedge \epsilon_3  \wedge 
\epsilon_4 \wedge k_3  \right)^2\\
&+ \left( \epsilon_1 \wedge \epsilon_2 \wedge \epsilon_3  \wedge 
\epsilon_4 \wedge k_4 \right)^2+ \text{$S_3$ permutations},\\
S_{\partial^0}=&\left( \epsilon_1 \wedge \epsilon_2 \wedge \epsilon_3  \wedge 
\epsilon_4\right)^2 +\text{$S_3$ permutations}.\\
\end{split}
\end{equation}

\begin{table}
\begin{center}
\begin{tabular}{ |c|c|c|c|}
\hline
 $\partial^n$ &${\bf 1_S}$ &${\bf 2_M}$ & ${\bf 1_A}$ \\
 \hline 
 $n=8$ &6 & 9 & 3 \\
 \hline
 $n=6$ &11 & 21 & 10 \\
 \hline
 $n=4$ &15 & 25 & 10 \\
 \hline
 $n=2$ &6 & 10 & 4 \\
 \hline
 $n=0$ &2 & 2 & 0 \\
 \hline    
\end{tabular}
\end{center}
\caption{$S_3$ classification of spin two quasi-invariant modules (primaries) for $D\geq 8$ organised by derivatives, where $n$ denotes the derivative order of the structures.}\label{tableqinvspin2}
\end{table}

\subsection{Scattering in D=4}\label{4Dsct}

 In lower dimensions we have additional parity-violating structures which can contribute to tree-level scattering \cite{Kravchuk:2016qvl, Chowdhury:2019kaq} but we will not attempt to classify them in this work. We now specialise to scattering in $D=4$ and restrict our analysis to the parity even sector. In $D=4$, $\polo_i$ is a one dimensional vector parametrized by a single number. The algebraic implication of this fact is that the inner products of the form $\polo_i \cdot \polo_j$ reduce to products of numbers. As a consequence the following structure which would have transformed under a ${\bf 3}$ now transforms in a ${\bf 1_S}$

\begin{equation}
    \polo_1 \cdot \polo_2 \polo_3 \cdot \polo_4 \sim \polo_1 \polo_2 \polo_3 \polo_4.
\end{equation}

The space of orthonormal vectors is now three dimensional and generated by $(\epsilon^L_i, \epsilon^T_i, \polo_i)$. We use rotational invariance in our problem to express our scattering amplitude in the rotated orthonormal frame $({\epsilon'}^L_i, {\epsilon'}^T_i, {\polo_i}')$, where,
\begin{equation}\label{d4polparametl1}
    \begin{split}
          {\polo_i}' &= \sin(\theta_i)\left ( \cos(\phi_i)\epsilon^L_i + \sin(\phi_i)\epsilon^T_i\right) + \cos(\theta_i) \polo_i,\\
        {\epsilon^{L}_i}^{\prime} &= \cos(\theta_i)\left ( \cos(\phi_i)\epsilon^L_i +  \sin(\phi_i)\epsilon^T_i\right) - \sin(\theta_i) \polo_i,\\
        {\epsilon^{T}_i}^{\prime} &= \left ( \cos(\phi_i)\epsilon^T_i - \sin(\phi_i)\epsilon^L_i\right),\\
    \end{split}
\end{equation}
where we have $\polo_i\cdot \polo_i=1$. Note that the condition $\epsilon_i\cdot k_i=0$ holds in the rotated frame as well. For spin two scattering, we also have the constraint of tracelessness. The complex basis of in-plane polarisations in the rotated frame can be written as 
\begin{equation}\label{d4polparamet}
    \begin{split}
       \epsilon^{\pm}_{i} (\theta_i, \phi_i) = & \sqrt{\frac{1}{2}} \left( {\epsilon^{L}_i}^{\prime} \pm  i {\epsilon^{T}_i}^{\prime}  \right)\\ =  & \left( \cos(\theta_i)\cos(\phi_i) \mp i \sin(\phi_i)\right)\epsilon^L_i + \left( \cos(\theta_i)\sin(\phi_i) \pm i \cos(\phi_i)\right)\epsilon^T_i - \sin(\theta_i)\polo_i.
    \end{split}
\end{equation}
 Now we can see that, $\epsilon_i^{+}(\pi - \theta_i, \phi_i) = -\epsilon_i^{-}(\theta_i, \phi_i )$. The in-plane polarisations in this rotated frame, i.e $\epsilon_i^{+}(\theta_i, \phi_i)$ suffices for a general spin two analysis in $D=4$, if we consider the $\theta_i$'s and $\phi_i$'s are arbitrary variables with, $0 \leq \theta_i  \leq \pi, ~~ 0\leq \phi_i < 2 \pi$ \footnote{It will be interesting to map this parametrization in $D=4$ to other ways of parametrizing massive spinning polarisations in the literature \cite{Arkani-Hamed:2017jhn}.}. 

For identical scattering with spin one external particles, there is a reduction in the number of linearly independent parity even quasi- invariant structures in $D=4$ compared to $D\geq 8$ that we have listed in this work (see eqs (4.52) and (4.53 ) of \cite{Kravchuk:2016qvl}). However, the Regge allowed S-matrices continue to be generated by the Lagrangians in eq \eqref{spin1crgalwdlag}. We have checked their linear independence using explicit parametrisation of the polarisation tensors in $D=4$ (using eqs \eqref{spinonepol} and \eqref{d4polparametl1} and the fact that $ \polo_i \cdot \polo_j \sim \polo_i \polo_j$ in $D=4$).

For spin two massive external states, the situation is a bit more dramatic, the amplitude generated by the six derivative term in eq \eqref{basisspin2id}, vanishes since it is topological in $D\leq 6$. This is easy to see from the structure of the S-matrix and the decomposition of the polarisation tensor we have implemented in this work. A generic polarisation $\epsilon^i$ corresponding to a spinning particle of mass $m_i$, is decomposed into a in-plane and out-of plane part 
\begin{eqnarray}
    \epsilon^i={\polo}^i + {\polp}^i,\qquad {\polp}^i = \alpha^i k^1 + \beta^i k^2 + \gamma^i k^3,    
\end{eqnarray}
where in $D=4$, ${\polo}^i$ is now just a number- more precisely it is constrained to be $\pm i$ if we normalise our polarisation tensors to be $|\epsilon^i|^2=1$ (recall $\epsilon^i\cdot \epsilon^i=0$ because of tracelessness). The six derivative S-matrix involves a seven dimensional $\varepsilon$-tensor and becomes immediately clear from the structure 

$$\varepsilon^{abcdefg}\epsilon^1_a\epsilon^2_b\epsilon^3_c\epsilon^4_d p^1_e p^2_f p^3_g$$
that we need ${\polo}^i$ to be atleast four-dimensional for this quantity to be non-zero (We have used momentum conservation to discard the fourth momenta $k^4$ throughout). Note that this argument is also true for the four and two derivative structure where we need ${\polo}^i$ to be atleast three and two dimensional respectively for the amplitude to be non-zero. In summary, for $D=4$, there is only one  parity even Regge allowed structure, generated by the 0 derivative Lagrangian in eq \eqref{basisspin2id}.

\subsection{CRG analysis of Massive Gravity} \label{CRGdRGT}
In this subsection we show that tree-level massive graviton scattering violate CRG in dRGT gravity \cite{deRham:2010ik, deRham:2010kj}. The effective Lagrangian is described by 

\begin{equation}
    \begin{split}
        S=\int d^4 x \sqrt{-g}\left( \frac{m_{P}^2}{2} R- \frac{m_{P}^2 m^2}{8} V(g,h)\right),
    \end{split}
\end{equation}
  where $R$ is the Riemann scalar built out of $g_{\mu\nu}=g^0_{\mu\nu} + h_{\mu\nu}$. Here $g^0_{\mu\nu}$ is the Minkowski metric $\eta_{\mu\nu}$, $h_{\mu\nu}$ is the massive spin two field, $m$ is its mass and $m_P$ is the reduced Planck mass. The interactions are concisely packaged into $V(g,h)= V_2(g,h) + V_3(g,h)+V_4(g,h)+\cdots$ as 
\begin{equation}
    \begin{split}
        V_2(g,h)&= b_1\langle h^2\rangle +b_2\langle h\rangle^2,\\
        V_3(g,h)&= c_1\langle h^3\rangle +c_2\langle h^2\rangle \langle h\rangle + c_3 \langle  h\rangle^3,\\
        V_4(g,h)&= d_1\langle h^4\rangle +d_2\langle h^3\rangle \langle h\rangle + d_3 \langle  h^2\rangle^2 +d_4 \langle  h^2\rangle\langle  h\rangle^2+ d_5 \langle  h\rangle^4,\\
    \end{split}
\end{equation}
where $\langle h\rangle= h_{\mu \nu} g^{\mu \nu},~ \langle h^2\rangle= h_{\mu \nu} g^{\mu \alpha}h_{\alpha \beta} g^{\beta \nu}$ and so on. The ghost free conditions yields the following relations, 
\begin{equation}\label{drgt}
    \begin{split}
        b_1&=-b_2=1,\qquad c_1=2c_3+\frac{1}2, \qquad c_2=-3c_3-\frac{1}{2},\qquad d_1= -6d_5+\frac{3 c_3}{2}+\frac{5}{16}\\
        d_2&=8d_5-\frac{3 c_3}{2}-\frac{1}{4},\qquad d_3=3d_5-\frac{3c_3}{4}-\frac{1}{16},\qquad d_4=-6d_5+\frac{3 c_3}{4}.
    \end{split}
\end{equation}

We wish to compute exchange diagrams and contact terms which contribute to tree-level scattering of $2 \rightarrow 2$ massive spin two fields in this theory and study their Regge growth. Our Regge limit probes the energies, 
\begin{equation}\label{energydrgt}
    m^2\ll s\ll M^2,\qquad M\sim (m^3 m_P)^\frac{1}{4}, 
\end{equation}
such that loops are suppressed and we can impose CRG on the tree-level amplitude. We follow \cite{Hinterbichler:2011tt} and use $g_{\mu\nu}= \eta_{\mu\nu}+h_{\mu\nu}$ to expand this Lagrangian to cubic and quartic in order to get the vertices. We use the following quadratic expansions of the inverse metric and its determinant,  
\begin{eqnarray}
    g^{\mu\nu}=\eta^{\mu\nu} - h^{\mu\nu}+ h^{\mu}_\alpha h^{\alpha \nu}+ O(h^3),\qquad \sqrt{-g}=\left( 1 +\frac{1}{2} h^\mu_\mu -\frac{1}{4} \left(h^{\mu\nu} h_{\mu\nu}-\frac{{h^\mu_\mu}^2}{2}\right)\right)+ O(h^3).\nonumber\\
\end{eqnarray}
At quadratic order, the Lagrangian in eq \eqref{drgt} gives rise to the following ghost-free spin two propagator \footnote{Note the change in normalisation of the propagator in comparison with eq 2.44 of \cite{Hinterbichler:2011tt}.}.  
\begin{equation}\label{massivepropagator}
   \begin{split}
P_{\mu_{1}\mu_{2} , \nu_{1}\nu_{2}} (p)&= \frac{4}{m_P^2}\frac{1}{p^2 + m^2}\Bigg(\frac{1}{2}\Big(\theta_{\mu_{1}\nu_{1}}\theta_{\mu_{2}\nu_{2}}+\theta_{\mu_{1}\nu_{2}}\theta_{\mu_{2}\nu_{1}}\Big) - \frac{1}{3}\theta_{\mu_{1}\mu_{2}}\theta_{\nu_{1}\nu_{2}}\Bigg),\\
\theta_{ab}&= \eta_{ab}+\frac{p_a p_b}{m^2}.
   \end{split}
\end{equation}
It was shown in \cite{Park:2010rp, Park:2010zw, deRham:2013qqa}, that despite the higher derivative terms in the propagator, the loop contributions are suppressed in $m_P$ (and hence $M$) and we can safely consider tree-level amplitudes in the energy range in eq \eqref{energydrgt}. For computing the tree-level, scattering 
relevant three point vertices and four point vertices have to be evaluated. We note that the off-shell cubic and the on-shell quartic vertices from $\sqrt{-g} V(g,h)$ are given by

\begin{equation}
\begin{split}
    g^{(3)}_V&= \frac{-m^2m_P^2}{8} \left(b_1 \left(\frac{h_{\alpha\beta}h_{\beta\alpha} h^\mu_\mu}{2} -2 h_{\alpha\beta}h_{\beta\gamma} h_{\gamma\alpha}\right) + b_2 \left(\frac{ ({h^\mu_\mu})^3}{2}-2h_{\alpha\beta}h_{\beta\alpha} h^\mu_\mu\right) \right.\\
        &\left. +c_1 h_{\alpha\beta}h_{\beta \gamma} h_{\gamma\alpha} + c_2 h_{\alpha\beta}h_{\beta\alpha} h^\mu_\mu +c_3 (h^\mu_\mu)^3\right),\label{3ptV}\\
         g^{(4)}_V &= \frac{-m^2m_P^2}{8} \left((3b_1-3c_1+d_1) (h_{\alpha\beta}h_{\beta\gamma} h_{\gamma\mu} h_{\mu\alpha})+ (-\frac{b_1}{4}+b_2-c_2+d_3) (h_{\alpha\beta}h_{\beta \alpha})^2 \right), 
        \end{split}
         \end{equation}

In these expansions, the indices of $h_{\mu\nu}$ are raised and lowered by the flat space background metric. We have also imposed $h_\mu^\mu=0$ in deriving the on-shell four point couplings. The relevant couplings from the  $\sqrt{-g} R$ are given by,

\begin{equation}
    \begin{split}
        g^{(3)}_R &=\frac{m_P^2}{2}
        \left(\frac{1}{8} (h^\mu_\mu)^2 \partial _{\sigma }\partial _{\rho }h_{\rho \sigma }-\frac{1}{4} h_{\mu \nu }^2 \partial _{\sigma }\partial _{\rho }h_{\rho \sigma }-\frac{3}{4} h_{\mu \nu } \partial _{\mu }h_{\rho \sigma } \partial _{\nu }h_{\rho \sigma }+h_{\mu \nu } \partial _{\sigma }h_{\mu \rho } \partial _{\nu }h_{\rho \sigma }+\frac{1}{2} h_{\mu \nu } \partial _{\sigma }h_{\mu \rho } \partial _{\rho }h_{\nu \sigma }\right.\\&\left.-\frac{3}{2} h_{\mu \nu } \partial _{\sigma }h_{\mu \rho } \partial _{\sigma }h_{\nu \rho }+h_{\mu \nu } \partial _{\rho }h_{\mu \rho } \partial _{\sigma }h_{\nu \sigma }+2 h_{\mu \nu } \partial _{\sigma }h_{\rho \sigma } \partial _{\nu }h_{\mu \rho }-h_{\mu \nu } \partial _{\sigma }h_{\rho \sigma } \partial _{\rho }h_{\mu \nu }+h_{\mu \nu } h_{\mu \rho } \partial _{\rho }\partial _{\sigma }h_{\nu \sigma }\right.\\&\left.+h_{\mu \nu } h_{\mu \rho } \partial _{\sigma }\partial _{\rho }h_{\nu \sigma }-h_{\mu \nu } h_{\mu \rho } \partial _{\sigma }\partial _{\sigma }h_{\nu \rho }+h_{\mu \nu } h_{\rho \sigma } \partial _{\sigma }\partial _{\nu }h_{\mu \rho }-h_{\mu \nu } h_{\rho \sigma } \partial _{\sigma }\partial _{\rho }h_{\mu \nu }-\partial _{\rho }h^\delta_\delta h_{\mu \nu } \partial _{\nu }h_{\mu \rho }\right.\\&\left.+\frac{1}{2} \partial _{\rho }h^\delta_\delta h_{\mu \nu } \partial _{\rho }h_{\mu \nu }-\partial _{\nu }h^\delta_\delta h_{\mu \nu } \partial _{\rho }h_{\mu \rho }-h_{\mu \nu } h_{\mu \rho } \partial _{\rho }\partial _{\nu }h^\delta_\delta+\frac{1}{4} \partial _{\sigma }\partial _{\sigma }h^\delta_\delta h_{\mu \nu }^2+\frac{1}{4} \partial _{\mu }h^\delta_\delta \partial _{\nu }h^\delta_\delta h_{\mu \nu }+\frac{3}{8} h^\delta_\delta \left(\partial _{\sigma }h_{\nu \rho }\right){}^2\right.\\&\left.-\frac{1}{4} h^\delta_\delta \partial _{\rho }h_{\nu \sigma } \partial _{\sigma }h_{\nu \rho }-\frac{1}{2} h^\delta_\delta \partial _{\nu }h_{\nu \rho } \partial _{\sigma }h_{\rho \sigma }-\frac{1}{2} h^\delta_\delta h_{\nu \rho } \partial _{\rho }\partial _{\sigma }h_{\nu \sigma }-\frac{1}{2} h^\delta_\delta h_{\nu \rho } \partial _{\sigma }\partial _{\rho }h_{\nu \sigma }+\frac{1}{2} h^\delta_\delta h_{\nu \rho } \partial _{\sigma }\partial _{\sigma }h_{\nu \rho }\right.\\&\left.+\frac{1}{2} h^\delta_\delta h_{\nu \rho } \partial _{\rho }\partial _{\nu } h^\delta_\delta+\frac{1}{2} h^\delta_\delta \partial _{\rho } h^\delta_\delta \partial _{\sigma }h_{\rho \sigma }-\frac{1}{8} (h^\mu_\mu)^2 \partial _{\sigma }\partial _{\sigma }h^\delta_\delta-\frac{1}{8} h^\delta_\delta \left(\partial _{\rho }h^\delta_\delta\right){}^2 \right), \\
        g^{(4)}_R &= \frac{m_P^2}{2}
        \left(-\frac{3}{16} \left(\partial _{\alpha }h_{\rho\sigma }\right){}^2 h_{\mu\nu }^2+\frac{1}{8} \partial _{\sigma }h_{\rho\alpha } \partial _{\alpha }h_{\rho\sigma } h_{\mu\nu }^2+\frac{1}{4} \partial _{\rho }h_{\rho\sigma } \partial _{\alpha }h_{\sigma\alpha } h_{\mu\nu }^2+\frac{3}{4} \partial _{\nu }h_{\sigma\alpha } \partial _{\rho }h_{\sigma\alpha } h_{\mu\nu } h_{\mu\rho }\right.\\&\left.-\partial _{\rho }h_{\sigma\alpha } \partial _{\alpha }h_{\nu\sigma } h_{\mu\nu } h_{\mu\rho }-\frac{1}{2} \partial _{\sigma }h_{\rho\alpha } \partial _{\alpha }h_{\nu\sigma } h_{\mu\nu } h_{\mu\rho }+\frac{3}{2} \partial _{\alpha }h_{\nu\sigma } \partial _{\alpha }h_{\rho\sigma } h_{\mu\nu } h_{\mu\rho }-\partial _{\sigma }h_{\nu\sigma } \partial _{\alpha }h_{\rho\alpha } h_{\mu\nu } h_{\mu\rho }\right.\\&\left.-2 \partial _{\rho }h_{\nu\sigma } \partial _{\alpha }h_{\sigma\alpha } h_{\mu\nu } h_{\mu\rho }+\partial _{\sigma }h_{\nu\rho } \partial _{\alpha }h_{\sigma\alpha } h_{\mu\nu } h_{\mu\rho }+\frac{1}{6} \partial _{\alpha }\partial _{\sigma }h_{\sigma\alpha } h_{\mu\nu } h_{\mu\rho } h_{\nu\rho }-\partial _{\sigma }\partial _{\alpha }h_{\rho\alpha } h_{\mu\nu } h_{\mu\rho } h_{\nu\sigma }\right.\\&\left.-\partial _{\alpha }\partial _{\sigma }h_{\rho\alpha } h_{\mu\nu } h_{\mu\rho } h_{\nu\sigma }+\partial _{\alpha }\partial _{\alpha }h_{\rho\sigma } h_{\mu\nu } h_{\mu\rho } h_{\nu\sigma }+\frac{1}{4} \partial _{\sigma }\partial _{\alpha }h_{\rho\alpha } h_{\mu\nu }^2 h_{\rho\sigma }+\frac{1}{4} \partial _{\alpha }\partial _{\sigma }h_{\rho\alpha } h_{\mu\nu }^2 h_{\rho\sigma }\right.\\&\left.-\frac{1}{4} \partial _{\alpha }\partial _{\alpha }h_{\rho\sigma } h_{\mu\nu }^2 h_{\rho\sigma }-\frac{1}{2} \partial _{\nu }h_{\sigma\alpha } \partial _{\rho }h_{\mu\alpha } h_{\mu\nu } h_{\rho\sigma }+\frac{3}{2} \partial _{\rho }h_{\mu\alpha } \partial _{\sigma }h_{\nu\alpha } h_{\mu\nu } h_{\rho\sigma }\right.\\&\left.-\partial _{\nu }h_{\mu\alpha } \partial _{\sigma }h_{\rho\alpha } h_{\mu\nu } h_{\rho\sigma }+\partial _{\sigma }h_{\rho\alpha } \partial _{\alpha }h_{\mu\nu } h_{\mu\nu } h_{\rho\sigma }-\partial _{\sigma }h_{\nu\alpha } \partial _{\alpha }h_{\mu\rho } h_{\mu\nu } h_{\rho\sigma }+\frac{3}{4} \partial _{\alpha }h_{\mu\rho } \partial _{\alpha }h_{\nu\sigma } h_{\mu\nu } h_{\rho\sigma }\right.\\&\left.-\frac{1}{4} \partial _{\alpha }h_{\mu\nu } \partial _{\alpha }h_{\rho\sigma } h_{\mu\nu } h_{\rho\sigma }-2 \partial _{\nu }h_{\mu\rho } \partial _{\alpha }h_{\sigma\alpha } h_{\mu\nu } h_{\rho\sigma }+\partial _{\rho }h_{\mu\nu } \partial _{\alpha }h_{\sigma\alpha } h_{\mu\nu } h_{\rho\sigma }+\partial _{\rho }\partial _{\nu }h_{\sigma\alpha } h_{\mu\nu } h_{\mu\rho } h_{\sigma\alpha }\right.\\& \left.-\partial _{\rho }\partial _{\alpha }h_{\nu\sigma } h_{\mu\nu } h_{\mu\rho } h_{\sigma\alpha }-\partial _{\alpha }\partial _{\rho }h_{\nu\sigma } h_{\mu\nu } h_{\mu\rho } h_{\sigma\alpha }+\partial _{\alpha }\partial _{\sigma }h_{\nu\rho } h_{\mu\nu } h_{\mu\rho } h_{\sigma\alpha } \right), 
    \end{split}
\end{equation}

where we have obtained the off shell three point and the on-shell four point vertices from the $\sqrt{-g} R$ using xAct tensor programme \cite{DBLP:journals/corr/abs-0803-0862}. The four point scattering therefore consists of exchange diagrams which are generated by two vertices $g^{(3)}_V$ and $g^{(3)}_R$ and contact diagrams generated by $g^{(4)}_V$ and $g^{(4)}_R$. The exchange diagram  is constructed by stitching together two three point functions as explained in section 7.2.2 of \cite{Chowdhury:2019kaq} using the massive spin two propagator in eq \eqref{massivepropagator}. For this purpose, it is necessary that we obtain the off-shell three point vertices (i.e without imposing the EOM \eqref{eomspin2}). Thus we have three  different kinds of exchange diagrams corresponding to three different choices of the vertices.
\begin{equation}
    A_{g^{(3)}_V-g^{(3)}_V},\qquad A_{g^{(3)}_V-g^{(3)}_R}, \qquad A_{g^{(3)}_R-g^{(3)}_R},
\end{equation}
The computation of these diagrams can be automated. We also have contact term amplitudes 
\begin{equation}
    A_{g^{(4)}_V},\qquad A_{g^{(4)}_R}.
\end{equation}
After a tedious calculation the complete four point amplitude is given by the sum of each individual pieces \footnote{As a consistency check we reproduced  tree-level scattering amplitude in Einstein gravity from $g^{(4)}_R$ and $g^{(3)}_R$ using this procedure but with a massless spin two propagator.}. The final answer is quite complicated and a \texttt{Mathematica} file containing the answer will be available on request. Using our explicit polarisation parametrisation in eqs \eqref{polpparaspinone} and \eqref{eomspin2}, we find that the Regge limit depends on  the tunable dimensionless coupling $c_3$ and the amplitude grows like $O(s^3)$.  The coefficient $d_5$ in the amplitude precisely occurs in the combination given by the zero derivative CRG allowed Lagrangian in eq \eqref{basisspin2id} and hence drops out in the Regge limit. However there exists a particular value of $c_3$ for which the leading behaviour of the four point amplitude can be ameliorated. 
\begin{equation}
c_3=\frac{1}{4}.
\end{equation}

Thus dRGT massive gravity violates CRG for all other possible values of couplings $c_3$ and $d_5$. This is also the conclusion of a causality based analysis by \cite{Camanho:2016opx}. Authors of  \cite{Haring:2022cyf}, showed that for tree-level amplitudes, local growth of S-matrices (i.e growth for energy scales below the cutoff but larger than other energy scales in the problem) is bounded by $s^2$ at large $s$ and fixed $t$, as a consequence of the unitarity of the full quantum scattering amplitude. If a theory violates CRG, which is a tree-level criteria, it might be sick when one considers the unitarity of the full S-matrix as well (as shown in \cite{Bellazzini:2023nqj} using dispersion relations). In summary, it seems to be crucial that while constructing such theories of massive gravity, CRG is an important criteria to impose if we want to construct massive theories of gravity. 

Our analysis involving contact terms in this paper shows promise in this regard. One might approach the problem in a bootstrap sense. Let us restrict to two derivative theories of gravity (say) and classify all amplitudes upto possible two derivative for three point amplitude and upto six derivative for four point amplitudes. The reason for considering upto six derivatives in contact S-matrices is that we might generate such six derivative amplitude when computing exchange diagrams involving two derivative three point Lagrangian couplings in our proposed theory (due to the higher derivative terms in the propagator eq \eqref{massivepropagator}). We now consider the space of $2 \rightarrow 2$ scattering amplitudes generated by these three point and four point amplitudes. Since we want ghost-free theories of gravity, we use the massive spin two propagator derived from ghost-free gravity in eq \eqref{massivepropagator} to construct the exchange diagrams. One then demands that arbitrary linear combination of these amplitudes cannot grow faster than $s^2$ at large $s$ and fixed $t$. It will be interesting to find out if there exists a non-zero solution for this problem other than dRGT massive gravity theory. One can then try to see if these other CRG allowed massive gravity theories obey the full UV consistency derived through dispersion relations.

\subsection{Ambiguity in Inversion formula}\label{inversionAmb}

In this subsection we interpret the CRG allowed Lagrangian structures as ambiguity in inversion formula in large $N$ conformal field theories (\cite{Caron-Huot:2017vep, Simmons-Duffin:2017nub, Turiaci:2018dht}). Let us review the argument of \cite{Turiaci:2018dht} for ambiguity in the inversion formula for scalars. Four point functions of primary scalar operators in conformal field theories are constrained by symmetry upto a function of crossratios $u$ and $v$, which we define below. Operator product expansion enables us to express this as a series expansion involving a purely kinematic piece called the conformal block and a part containing information about the CFT.   

\begin{equation}
\begin{split}
\langle O(x_1) O(x_2) O(x_3) O(x_4)\rangle &= \frac{1}{x_{12}^{\Delta_O}}\frac{1}{x_{34}^{\Delta_O}} {\mathfrak g}(u,v)= \frac{1}{x_{12}^{\Delta_O}}\frac{1}{x_{34}^{\Delta_O}} \sum_{\Delta, J} C_{\Delta, J} G_{\Delta, J} (u,v),\\
\text{where}&,\qquad
 u=z \bar{z}= \frac{x_{12}^2 x_{34}^2}{x_{13}^2x_{24}^2},\qquad v=(1-z)(1-\bar{z})= \frac{x_{23}^2 x_{14}^2}{x_{13}^2x_{24}^2},
\end{split}
\end{equation}

where $C_{\Delta, J}$ is the product of OPE coefficients, $G_{\Delta, J} (u,v)$ is the conformal block and the sum over $\Delta$ and $J$ runs over the dimensions and spin of the primary operators that appear in the OPE expansion of two external operators. The data of the CFT is  completely specified by the scaling dimensions and spin of  the primary operators and their three point functions $C_{\Delta, J}$. In \cite{Caron-Huot:2017vep, Simmons-Duffin:2017nub}, it was shown that this data can be obtained from the double discontinuities of the correlator. Schematically this was expressed in \cite{Turiaci:2018dht} in the following manner,  

\begin{equation}
\begin{split}
I_{\Delta ,J} &= \int_0^1 dz d \bar{z} K^t_{\Delta, J}(z, \bar{z}) \langle [O(x_2),O(x_3)][O(x_1),O(x_4)]\rangle \\
&+ (-1)^J \int^0_{-\infty} dz d \bar{z}K^t_{\Delta, J}(z, \bar{z}) \langle [O(x_2),O(x_4)][O(x_1),O(x_3)]\rangle,
\end{split}
\end{equation}
where the functions $K^t_{\Delta, J}(z, \bar{z})$ are explicitly known.      
The function $I_{\Delta ,J}$  encodes the data for the CFT. The poles in $\Delta $ correspond to the dimensions of the operators appearing in the OPE of the two external operators and the residue at these poles generate the OPE coefficients.
The RHS of the formula involves double discontinuity of the correlator which receives non-zero contribution from all the single trace operators in the spectrum but is unaffected by double trace operators. Hence the inversion formula enables us to extract double trace data about the CFT from just knowing about the single trace operators. The residues are analytic in spin but for a generic CFT this formula valid only for $J>1$. This bound follows from the boundedness of the CFT correlator in Regge limit, a kinematic limit of the Correlator in the Lorentzian regime \cite{Caron-Huot:2017vep}. This is particularly interesting for large $N$ CFTs where this constraint is weaker \cite{Maldacena:2015waa} and we can get data only for $J>2$. Hence, two correlators having the same double discontinuities can differ by presence of $J=0$ and $J=2$ double trace operators in the spectrum. On the other hand, in large $N$ CFTs, the double trace operators appearing in the large $N$ crossing equations of the CFT four point function are in one to one correspondence with local counterterms in the bulk \cite{Heemskerk:2009pn}.  Putting these two pieces of argument together, in \cite{Turiaci:2018dht}, the authors proved a theorem that the four point functions of scalar operators in large $N$ CFTs are completely fixed by the double discontinuity of the correlator and three local counterterms in $AdS$ with arbitrary coefficients.\footnote{Similar arguments have been used in \cite{Silva:2021ece} to constrain the mixed four point functions in large $N$ Chern Simons theories with vector matter.} These set of counter terms in $AdS$ give rise to CFT correlators whose Regge growth are bounded by Chaos bound \cite{Maldacena:2015waa}. In particular, the correlators generated by bulk contact like terms in the Regge limit behave as \cite{Chandorkar:2021viw},
\begin{equation}
\begin{split}
\lim_{\sigma \rightarrow 0}~{\mathfrak g}^{CR}(z,\bar{z}) \propto \frac{h^{CR}(\rho)}{\sigma^{A'-1}}, \qquad z= \sigma e^{-\rho},~~ \bar{z}= \sigma e^{\rho},
\end{split}
\end{equation}
where the superscript $CR$ denotes that limit is taken in a Lorentzian Regge configuration, which is obtained by analytically continuing from the euclidean correlation function by an anticlockwise rotation of the cross ratio $\bar{z}$ about the point 1, $(1-\bar{z})\rightarrow (1-\bar{z}) e^{-2\pi i}$ while $z$ is held fixed in the upper half plane (UHP). Causally, this the configuration in which $O(x_2)$ is in the future light cone of $O(x_3)$ and $O(x_4)$ is in the future light cone of $O(x_1)$ but these pairs of particles are space like separated from one another. Regge limit corresponds to boosting the operators $O(x_2)$ and $O(x_4)$ in this Lorentzian Regge configuration, which translates to $z \rightarrow 0, \bar{z} \rightarrow 0$ with $z/\bar{z}$ held fixed (also expressed as $\sigma \rightarrow 0$ at fixed $\rho$). For large $N$ CFTs, Chaos bound implies $A'\leq 2$.

There exists a different Lorentzian kinematic limit for correlators of large $N$ CFTs called the ``Bulk point" limit which generates tree-level flat space scattering amplitudes corresponding to bulk interactions \cite{Maldacena:2015iua,  Penedones:2010ue, Gary:2009ae}. For a Lagrangian which generates flat space S-matrix which grows in Regge limit as $s^A$, the small $\sigma$ behaviour of the bulk point correlator takes the following form 
\begin{equation}\label{bp}
\begin{split}
\lim_{\sigma \rightarrow 0,~\rho \rightarrow 0 }~{\mathfrak g}^{CS}(u,v) \propto \frac{1}{\sigma^{A-1} \rho^\alpha},
\end{split}
\end{equation}
where the superscript $CS$ denotes that limit is taken in a Lorentzian scattering configuration, which is analytically continued from the euclidean correlation function by an anticlockwise rotation of the cross ratio $\bar{z}$ about the point 1, $(1-\bar{z})\rightarrow (1-\bar{z}) e^{-2\pi i}$ while $z$ is held fixed in the UHP followed by anti-clockwise rotation of $z \rightarrow z e^{-2\pi i}$ while $\bar{z}$ is held fined in UHP. Causally, this the configuration in which both $O(x_2), O(x_4)$ are in the future light cone of $O(x_3)$ and $O(x_1)$. Bulk point corresponds to the limit $z/\bar{z} \rightarrow 1$ or $\rho \rightarrow 0$. The coefficient $\alpha$ encodes the derivative order of the bulk interaction. For CFT correlators generated by contact term like interactions, one can analytically continue from the causally Regge sheet to causally scattering sheet \cite{Chandorkar:2021viw}. 
\begin{equation}\label{rgalytic}
\frac{h^{CR}(\rho)}{\sigma^{A'-1}} \rightarrow \frac{h^{CS}(\rho)}{\sigma^{A'-1}}.
\end{equation}
In particular, this analytic continuation does not affect the exponent of $\sigma$ for contact term interactions in the bulk. Eq \eqref{bp} captures the behaviour of the CFT correlator when both the cross ratios are small while \eqref{rgalytic} is for small $\sigma$ but finite $\rho$. For correlators of scalars, massless spin one and spin two, this analytic continuation between two different Lorentzian kinematics and detailed analysis of two order of limits of cross-ratios imply that flat space Lagrangians that violate CRG necessarily violate Chaos bound, when we study the Regge limit of the correlators generated by these Lagrangians in $AdS$.
\begin{equation}\label{crgchaos}
A\leq A'\leq 2.
\end{equation}

 Violation of Chaos bound however does not imply violation of CRG. Although this leaves the possibility that there might be Lagrangians which are CRG allowed but violate Chaos bound in $AdS$, we do not expect this to happen  from the structure of $AdS$ integrals involved in the proof for scalar, massless spin one and spin two particles (see \cite{Chandorkar:2021viw} for detailed analysis in this regard). In fact, in all the examples explicitly studied in \cite{Chandorkar:2021viw}, $A'=A$. Indeed, the set of scalar $AdS$ counterterms classified by \cite{Turiaci:2018dht}, are same as the set of scalar counterterms that are allowed by CRG for four point flat space scattering of scalars \cite{Chowdhury:2019kaq}. This argument then generalises to inversion formula for conserved spinning external particles \cite{Kravchuk:2018htv,Karateev:2018oml,Caron-Huot:2021kjy}. The local four photon and four graviton counterterms that are CRG allowed were classified in \cite{Chowdhury:2019kaq} and they can be interpreted as the local counterterms which are $AdS$ ambiguities for the inversion formula for four point functions of conserved currents and stress tensors. 

We would like to interpret our CRG allowed massive spin one and massive spin two Lagrangians, in eqs \eqref{spin1crgalwdlag} and \eqref{basisspin2id}, as ambiguities of the inversion formula corresponding to four point functions of unconserved currents and symmetric traceless tensors. Indeed one can check that using spinning $AdS$ propagators \cite{Costa:2014kfa}, the tree-level $AdS$ correlators generated by our CRG allowed Lagrangians all obey Chaos bound. 
However, even though we believe this is the exhaustive list of Regge allowed counter terms, we are unable to provide a proof of this statement due to the following reason. For massive external states, ``Bulk point" limit is not the correct kinematic limit to probe flat space scattering (Appendix I of \cite{Chandorkar:2021viw}). Hence a naive extrapolation of the chain of arguments for massless scattering, that led us to establishing that CRG violation implies violation of Chaos bound, to massive scattering is not feasible.  In particular, this leaves open the possibility that there might exist CRG violating interactions in flat space which obey Chaos bound in $AdS$ for massive particles. While we believe this cannot be the case, it would be nice to establish a proof of this fact. Recently much progress have been made in trying to understand flat space scattering of massive external particles from $AdS$ correlators \cite{Komatsu:2020sag, vanRees:2022zmr, Hijano:2019qmi, Li:2021snj}. Using this machinery it would be interesting to derive the analogue of eq \eqref{crgchaos} for massive scattering, which would constitute a proof of the statement that violation of CRG for S matrices of massive particles implies violation of Chaos bound for correlators of unconserved spinning operators in CFT.

\section{Conclusion}
In this work, we have classified the space of contact S-matrices (analytic in polarisation tensors and momenta) that contribute to  $2 \rightarrow 2$ tree-level scattering of massive spin one and spin two particles which obey CRG criteria. Our results are for both identical as well as non-identical scattering of particles in $D \geq 8$. In the first part of this work, we constructed the space of all allowed S-matrices that can contribute to such tree-level processes. As was done in \cite{Chowdhury:2019kaq}, it is useful to think of this space as a module of polarisation tensors over the ring of polynomials of Mandelstam invariants. The classification then entails constructing the generator of these modules and together with the polynomial of Mandelstam invariants, these modules generate the most general S-matrix at any derivative order. Unlike the massless case, the local module generators for massive spinning particles are easy to construct. Since there is no gauge redundancy, the local generators need not only be polynomials of field strength (for spin one) or Riemann tensor (for spin two). The classified structures can also be organised in irreducible representations of $S_4$ when they correspond to scattering of identical particles. Our process have been summarised in \S \ref{ecosms1} and \S\ref{ecosms2} and we present the explicit structures in the ancillary \texttt{Mathematica} files . The counting of number of linearly independent structures contributing to the scattering can also be thought of constructing $SO(D-3)$ singlets out of representations of little group of massive spinning particles. We obtain exact agreement between the group theory counting and explicit construction of structures. Our construction, although done for $D\geq 8$, can easily generalised to lower dimensions as well where we expect reduction of parity even structures and appearance of additional parity violating structures due to the Levi-Civita tensors corresponding to each dimensions. In lower dimensions the modules might not be freely generated as well. It would be interesting to verify this through Hilbert series for the massive spin one and spin two fields \cite{Chowdhury:2019kaq, Henning:2017fpj,Aharony:2003sx, deMelloKoch:2017dgi, deMelloKoch:2018klm}.
For non-identical scattering, each of the primaries corresponds to a local Lagrangian while for identical scattering, the quasi-invariants correspond to a Lagrangian coupling. 

In the second part of this work, we study the Regge behaviour (large $s$ at fixed $t$) of these structures for both identical and non-identical scattering and classify the couplings consistent with Classical Regge Growth criteria. For non-identical scattering, a single coupling can contribute to different processes (obtained by permutations of the external states). We demand that amplitude generated by a coupling grows slower than or equal to $s^2$ in the Regge limit for all possible choices of polarisation tensors and each of the possible processes. We generalise this study for the case of identical scattering as well. Our method and results have been summarised in \S \ref{CRGALLWDl1} and \S\ref{CRGALLOWED}. For the identical scattering, we find that only a finite number of module generators and their descendants are CRG allowed and the corresponding local Lagrangians are given by  eq \eqref{spin1crgalwdlag} for the massive spin one and eq \eqref{basisspin2id} for the massive spin two scattering. We specialise to the case of $D=4$, where remarkably only a single parity even structure survives for massive spin two, given by the zero derivative Lagrangian in eq \eqref{basisspin2id}. 

We show that dRGT theory \cite{deRham:2010ik, deRham:2017xox} violates CRG except for the special choice of its parameter values $c_3=\frac{1}{4}$. We do so by explicitly computing the four point tree-level amplitude in this theory.  We propose that one should consider CRG as a criteria that a theory must satisfy when constructing tree-level massive gravity theories. In particular, our preliminary analysis involving contact-terms suggest that it would be worthwhile exploring this problem in the space of scattering amplitudes in a bootstrap manner by considering both exchange diagrams and contact terms together. 

We, then, interpret our CRG allowed Lagrangians as possible $AdS$ contact term ambiguities that relevant for inversion formula for large $N$ CFTs. We believe, our list of CRG allowed counterterms is exhaustive despite the fact that it assumes CRG violation in massive flat space scattering implies violation of Chaos bound in the bulk (since the existing proof of the statement in \cite{Chandorkar:2021viw} deals with massless particles). It would be interesting to prove this using recent developments in massive scattering in flat space physics from $AdS$ \cite{Komatsu:2020sag, vanRees:2022zmr, Hijano:2019qmi, Li:2021snj}.   

In \cite{Chowdhury:2019kaq}, a broad goal of constraining classical theories of gravity was proposed. One of the conjectures was that there exist exactly three classical gravitational S-matrices that are consistent with a set of physically motivated ‘low energy’ constraints- Einstein S-matrix generated generated by Einstein Hilbert action, the type II S-matrix and the Heterotic S-matrix generated by their classical truncations. This conjecture implied but is not implied by two more conjectures. The only consistent classical gravitational S-matrix whose exchange poles are bounded in spin is the Einstein S-matrix and the only consistent classical gravitational S-matrix with massless spin two pole is the Einstein S-matrix. For $D\leq 6$, the second and third conjectures were shown to be true for four point scattering in the sense that there exists no finite polynomial deformation of Einstein gravity consistent with CRG criteria \cite{Chowdhury:2019kaq}. However to address the first conjecture, we need to consider scattering of massive spins, mixed scattering as well as higher point scattering. 

Let us consider dimensional reduction of pure Einstein gravity in  from $R^{1,D-1}$ dimensions to $R^{1,D-2} \times S^1/\Z_2$ \cite{Hang:2021fmp}. The $D-1$ dimensional theory has real massive spin one, massive spin two  as well as massless spin two, massless spin one and scalar particles. The interactions generating tree-level amplitudes in the dimensionally reduced theory can be worked out by linearising the Einstein gravity action to quartic orders. The three point and four point couplings of the dimensionally reduced fields are infinite in number and are highly fine tuned with very specific numerical coefficients appearing in front of them. The tree-level scattering amplitudes in $D$ dimensional pure gravity theory saturates CRG. If one were to study scattering amplitudes in the $D-1$ dimensional theory, this would be a consequence of cancellations in the Regge growth due to fine tuning in the Kaluza Klein (KK) theory. Thus this is a nice toy model for making progress on conjecture I of \cite{Chowdhury:2019kaq}. In the context of this toy model, we can now ask the following question. We take a linear combination of all possible two derivative three point functions and contact terms of massive spin one, spin two and massless spin one and two particles in a $D-1$ dimensional theory and try to see if there exist a solution by demanding that the scattering amplitude, that this linear combination generates, must obey CRG. One would expect that if we try to, say, impose CRG over just the space of four point couplings of massive spin one without taking into account the exchanges, we should find that no finite combinations of such four point couplings should be allowed because of our toy example. But as we show in \S \ref{CRGALLWDl1} that there exists a finite solution to this bootstrap problem. Thus a preliminary analysis leads to a conclusion that CRG is not enough to completely fix the spectrum to KK spectrum in this approach. Despite this, it would be interesting to see how close we can constrain the spectrum to KK theories using CRG by considering both exchange amplitudes and purely contact amplitudes together\footnote{We thank Abhijit Gadde for discussions regarding this issue.}. 

We leave these threads for future endeavours.

\acknowledgments

We would like to thank Shiraz Minwalla for collaboration at various stages of the project and very useful discussions throught the duration of the project. We would also like to thank Giulia Isabella and Shiraz Minwalla for very useful comments on the manuscript. This work of SDC is supported by the NSF Grant No.\ PHY2014195 and  Kadanoff fellowship at the University of Chicago. The work of VK and AR was supported by the Infosys Endowment for the study of the Quantum Structure of Spacetime and by the J C Bose Fellowship JCB/2019/000052. The work of SK was supported in part by an ISF, center for excellence grant (grant number 2289/18), Simons Foundation grant 994296, Koshland Fellowship and TIFR for hospitality. VK and AR would also like to acknowledge their debt to the people of India for their steady support to the study of the basic sciences.

\appendix
\section{$S_4$ representation theory} \label{s3} 
In this appendix we review the group theory for discrete group $S_4$, relevant for main text. The discussion closely follows \S 2.3 of \cite{Chowdhury:2019kaq}. $S_4$ the discrete group that permutes four objects, which, in the context of this paper are the four scattering particles. $S_4$ has a normal abelian subgroup generated by $\Z_2 \times \Z_2$, the double transposition. This normal subgroup is generated by $(P_{12}P_{34}, P_{13}P_{24})$ (where $P_{ij}$ swaps $i\leftrightarrow j$) and has four elements 
\begin{equation}\label{Z2Z2el}
    \begin{split}
        \{1,P_{12}P_{34},P_{13}P_{24},P_{14}P_{23}\}.
    \end{split}
\end{equation} 
Since the Mandelstam invariants are invariant under this normal subgroup, it is convenient to impose the $S_4$ symmetry in two steps. Given a polynomial of polarisation tensors and Mandelstam invariants, we first impose $\Z_2 \times \Z_2$ and then impose the remnant permutation group given by the coset $S_3=\frac{S_4}{\Z_2 \times \Z_2}$. The remnant symmetry group is generated by 
\begin{equation}\label{S3el}
    \begin{split}
        \{1,P_{12},P_{13},P_{23},P_{123},P_{231}\}.
    \end{split}
\end{equation} 
The total action of $S_4$ therefore corresponds to first gauge fixing an $S_4$ element to the form $(ijk4)$ using $\Z_2 \times \Z_2$ and then the left action of $S_3$ on $(123)$. Using the above factorization we first project polynomials of momenta and polarisations to the  $\mathbb{Z}_2 \times\mathbb{Z}_2$ invariants by using suitable projectors and then construct the $S_3$  irreps of those S-matrices using the projectors corresponding to the $S_3$ irreps. The $\mathbb{Z}_2 \times\mathbb{Z}_2$ invariant structure is also known as the ``quasi-invariant S-matrix" following the terminology in \cite{Chowdhury:2019kaq}. The projector corresponding to quasi-invariant S-matrices is,
\begin{equation}\label{z2z2proj}
    \Pi_0 = \frac{1}{4}\left(1 + P_{12}P_{34}+P_{13}P_{24}+P_{14}P_{23}\right).
\end{equation} 

Now, we construct projectors corresponding to different irreps of $S_3$. There are three irreducible representations of $S_3$,  
\begin{equation}
    {\bf 1_S}={\tyng(3)},\qquad {\bf 2_M}={\tyng(2,1)},\qquad {\bf 1_A}={\tyng(1,1,1)}.
\end{equation} 

$\mathbf{1_S}$ is the complete symmetric, $\mathbf{1_A}$ is the complete anti-symmetric, and  $\mathbf{2_M}$ is the 2 dimensional mixed symmetric representation. The corresponding projectors are given below:
\begin{equation}\label{s3proj}
    \begin{split}
        &\Pi_{\mathbf{1_S}} = \frac{1}{6}\left(1 + P_{12} + P_{23} + P_{13} + P_{231} + P_{123}\right),\quad
        \Pi_{\mathbf{1_A}} = \frac{1}{6}\left(1 - P_{12} - P_{23} - P_{13} + P_{231} + P_{123}\right),\\&
        \Pi^{(1)}_{\mathbf{2_M}} = \frac{1 + P_{12}}{2} - \Pi_{\mathbf 1_S},\quad
        \Pi^{(2)}_{\mathbf{2_M}} = \frac{P_{23} + P_{123}}{2} - \Pi_{\mathbf{1_S}},\quad
        \Pi^{(3)}_{\mathbf{2_M}} = \frac{P_{13} + P_{231}}{2} - \Pi_{\mathbf{1_S}}.\\
    \end{split}
\end{equation}
One can easily check that $\Pi^{3}_{\mathbf{2_M}}$ is not independent. The projectors satisfy the identity,

\begin{equation}
    \Pi^{(1)}_{\mathbf{2_M}} + \Pi^{(2)}_{\mathbf{2_M}}+ \Pi^{(3)}_{\mathbf{2_M}} = 0,
\end{equation}
as $\mathbf{2_M}$ is a 2 dimensional representation. These are the final set of $S_4$ projectors that we will use in the orbit classification S-matrices. It is also instructional to record the action of $S_3$ on the space of polynomials of Mandelstam invariants subject to the condition $s+t+u=4m^2$. It is easy to convince oneself that the polynomials of $(s,t,u)$ transforming 
in various irreps of $S_3$ subject to the condition of 
$s+t+u=4m^2$ are given by, 
\begin{equation}\label{polydesc}
\begin{split}
&f_{{\bf 1_S}}(s,t,u)= m^a(s^2+t^2+u^2)^b (s t u)^c,\\
&f^{(1)}_{{\bf 2_M}}(s,t,u)= m^a(s^2+t^2+u^2)^b (s t u)^c \{(2s-t-u), (2s^2-t^2-u^2)\},\\
&f_{{\bf 1_A}}(s,t,u)= m^a(s^2+t^2+u^2)^b (s t u)^c \left(s^2t-t^2s+t^2u-u^2t+u^2s-s^2u\right).\\
\end{split} 
\end{equation}

\section{Explicit expressions for polarisation vectors}\label{eepv}
We present the explicit expressions for 
$\mathcal{N}_1^L, \mathcal{N}_1^T, \mathcal{C}_1, \mathcal{C}_2, \mathcal{T}_1, \mathcal{T}_2$ used in eq \eqref{polpparaspinone}. 
\begin{equation}\label{Normalisation}
    \begin{split}
          \mathcal{N}_1^{L} =& \sqrt{\frac{m_2^2 k_{14}^2+(s-2m^2) \left(m_4^2 k_{12}-k_{14} k_{24}\right)}{m_1^2 \left(k_{24}^2-4 m_2^2 m_4^2\right)+m_2^2 k_{14}^2+k_{12} \left(m_4^2 k_{12}-k_{14} k_{24}\right)}},\\
          \mathcal{N}_1^{T} &=\sqrt{\frac{ m_1^2 k_{12}^2 k_{14}^2}{-m_4^2 k_{12}^2-m_2^2 k_{14}^2+k_{14} k_{24} k_{12}}},\\ \mathcal{C}_1=&\frac{m_1^2 \left(k_{14} k_{24}-2 m_4^2 k_{12}\right)}{m_2^2 k_{14}^2+k_{12} \left(m_4^2 k_{12}-k_{14} k_{24}\right)},\qquad \mathcal{C}_2=\frac{m_1^2 \left(k_{12} k_{24}-2 m_2^2 k_{14}\right)}{m_2^2 k_{14}^2+k_{12} \left(m_4^2 k_{12}-k_{14} k_{24}\right)},\\ \mathcal{T}_1 = & \frac{1}{k_{12}},\qquad \mathcal{T}_2 =  \frac{1}{k_{14}},\qquad {\epsilon_1^{L}}^2={\epsilon_1^{T}}^2=1.\\
          \end{split}
\end{equation}
Here we define, $k_{ij} = -2k_i.k_j = s_{ij} - m_i^2  - m_j^2 $. $s_{ij}$ is symmetric in $i$ and $j$ and $s_{12} =s_{34} =s$, $s_{23} =s_{14} =u$, and $s_{13} =s_{24} =t$.

\providecommand{\href}[2]{#2}\begingroup\raggedright\endgroup


\end{document}